\documentclass[twocolumn,amsmath,nosuperscriptaddress,amssymb,nopacs,nofootinbib,prb]{revtex4-2}
\usepackage[usenames,dvips]{color}
\usepackage{graphicx}
\usepackage{dcolumn}
\usepackage{bm}
\usepackage{mathtools}
\usepackage{epstopdf}
\usepackage{esint}
\usepackage{xcolor}
\usepackage{dsfont}
\usepackage[outline]{contour}
\usepackage{nicefrac}
\usepackage{multirow}
\usepackage[colorlinks=true,citecolor=magenta,linkcolor=magenta,urlcolor=magenta]{hyperref}

\newcommand{\bea}{\begin{eqnarray}}
\newcommand{\eea}{\end{eqnarray}}
\newcommand{\bt}{\textbf}

\newcommand{\phd}{\phantom{\dag}}
\newcommand{\ph}{\phantom{.}}

\newcommand{\noi}{\noindent}
\newcommand{\no}{\nonumber}

\begin{document}
\def\v#1{{\bf #1}}

\title{Dynamical Chiral Symmetry and Symmetry-Class Conversion\\in Floquet Topological Insulators}

\author{Mohamed Assili}
\email{m.assili@itp.ac.cn}
\author{Panagiotis Kotetes}
\email{kotetes@itp.ac.cn}
\affiliation{CAS Key Laboratory of Theoretical Physics, Institute of Theoretical Physics, Chinese Academy of Sciences, Beijing 100190, China}

\vskip 1cm

\begin{abstract}
In this work, we discuss properties with no static counterpart arising in Floquet topological insulators with a dynamical chiral symmetry (DCS), i.e., a chiral symmetry which is present while driving. We explore the topological properties of Floquet insulators possessing a DCS which either does or does not survive upon taking the static limit. We consider the case of harmonic drives and employ a general framework using the quasi-energy operator in frequency space. We find that for a DCS with no static analog, the presence of driving has a negligible impact on the topological phases associated with zero quasi-energy. In stark contrast, topological gaps can open at $\pi$ quasi-energy and mainly occur at momenta where the driving perturbation vanishes. We confirm the above general predictions for an extended Kitaev chain model in the BDI symmetry class. Another possibility that opens up when adding the drive, while preserving chiral symmetry, is symmetry-class conversion. We demonstrate such an effect for a static CI class Hamiltonian which is topologically trivial in 1D. By considering a suitable driving, we obtain a CI$\rightarrow$AIII transition, which now enables the system to harbor topological $\pi$-modes. Notably, the arising topological phases strongly depend on whether the DCS has a static analog or not. Our results bring Floquet insulators with nonstandard DCS forward as ideal candidate platforms for engineering and manipulating topological $\pi$-modes.
\end{abstract}

\maketitle

\section{Introduction}

Periodically driven systems have recently attracted significant interest~\cite{Oka,Inoue}, since they exhibit distinct thermodynamic and, most remarkably, to\-po\-lo\-gi\-cal features which are absent in their static counterparts~\cite{kitagawa2010a,kitagawa2010b,lindner2011,Jiang,WMLiu,lindner2013,leon2013,DELiu2013,RudnerLevin,Ghosh2022}. A detailed review of the unique pro\-per\-ties that dictate these also-called Floquet systems can be found in Refs.~\onlinecite{CayssolRev,LuRev,BukovRev,CooperRev,OkaRev,RudnerRev,GuoRev,GhoshRev}. Up to date, there has been a long list of successful expe\-ri\-men\-tal realizations of Floquet platforms and in parti\-cu\-lar Floquet topological insulators (FTIs)~\cite{KitagawaExp,Rechtsman2013,Jotzu,JiminezGarcia,Peng,Maczewsky,Mukherjee,Hu2015,ChengPi,wintersperger2020}.

One of the hallmark properties of FTIs is the emergence of to\-po\-lo\-gi\-cal\-ly protected boun\-dary states, known as topological $\pi$-modes. There exists already a rich va\-rie\-ty of theoretical proposals for experimentally rea\-li\-zing and observing such modes. For instance, in one spatial dimension this can be achieved in driven systems which either emulate the Su-Schrieffer-Heeger (SSH)~\cite{SSH} or the Kitaev chain model~\cite{kitaev2001}. In the latter case, the ari\-sing topological $\pi$-modes are additionally charge neutral, hence constituting Majorana excitations. These are often referred to as Majorana $\pi$ modes (MPMs) in ana\-lo\-gy to the well studied Majorana zero modes (MZMs). A number of theoretical pro\-po\-sals have appeared with blueprints on how to engineer MZMs and MPMs in superfluids~\cite{Jiang,WMLiu,DELiu2013}, in p-wave superconductors~\cite{Tong2013,Thakurathi,Benito2014,li2017,li2018,wang2020,Wu2023,Roy2023arXiv}, and systems involving conventional superconductors~\cite{Reynoso,WMLiuHet,klinovaja2016,Thakurathi2017,Liu2019,YangZhesen,Cayao}. The possibility of the on-demand generation of MZMs and MPMs does not only enhance the functionalities of topological platforms but, most importantly, it enables carrying out effective braiding protocols in strictly 1D using a single topological nanowire~\cite{BomantaraPRL,bomantara2018,BelaBauer}.

The pursuit of inducing multiple Majorana excitations in a single segment in order to facilitate braiding can be further assisted by imposing chiral symmetry to the system~\cite{Tong2013,Wu2023}. In fact, chiral symmetry is an inhe\-rent ingredient of the particle-hole symmetric SSH model and the disorder-free Kitaev chain which, under these conditions, {\color{black}they belong to AIII and BDI symmetry classes~\cite{Comment}.} However, without fine tuning, these models end up in classes A and D. This reveals that chiral symmetry is crucial for achieving non\-tri\-vial topology in the SSH model, since class A is trivial in odd spatial dimensions~\cite{Ryu2010}. Notably, such a limitation does not appear for the Kitaev chain due to a built-in charge-conjugation symmetry~\cite{AZ}. Therefore, chiral symmetry is a vital ingredient for engineering nontrivial topology in 1D, as well as for creating multiple edge modes~\cite{Zhou2018,Zhou2020,Roy2023}, as mentioned above.

The exploration of Floquet systems with chiral symmetry, that we here term dynamical chiral symmetry (DCS), was motivated in early works by the investigation of topological quantum walks~\cite{kitagawa2010a,Obuse,Asboth2012, AsbothObuse,Tarasinski,Asboth2014}. Static FTIs with chiral symmetry are classified in terms of a win\-ding number~\cite{Ryu2010}, which predicts the number of topologically-protected modes per boundary. By extension, FTIs with DCS are characte\-ri\-zed by an additional winding number which establishes a bulk-boundary correspondence for $\pi$ modes. One of the possible methods to calculate these two topological invariants relies on evaluating the time-evolution ope\-ra\-tor (TEO) of the driven system at half a driving period~\cite{Asboth2014}. This time-domain-based approach has so far been extensively explored, since it is well-suited for deriving analytical results when consi\-de\-ring ``kicked" or stepwise evolving driving protocols. For harmonic drives instead, which are routinely employed in realistic experiments, the TEO approach becomes ana\-ly\-ti\-cal\-ly intractable and numerically inefficient. In such situations, it is advantageous to define topological inva\-riants in frequency space~\cite{RudnerLevin}. This approach relies on the so-called quasi-energy operator (QEO)~\cite{eckardt2015} and its systematic study has recently attracted attention~\cite{Kennes}.

In this Manuscript, we focus on exactly this category of harmonically-driven chiral-symmetric systems and further expand on the framework for inferring the Floquet to\-po\-lo\-gi\-cal invariants at quasi-energies $\varepsilon=\{0,\pi/T\}$. These are here expressed in terms of the driving period $T$, with the reduced Planck constant $\hbar$ set equal to unity. We construct the topological invariants by defining sui\-table winding numbers of the QEO in extended frequency space. In analogy to the static case, we first identify the general unitary transformation which brings the quasi-energy ope\-ra\-tor for $\varepsilon T=\{0,\pi\}$ to the so-called canonical basis, in which the QEO becomes block off-diagonal and the respective chiral-symmetry ope\-ra\-tor block dia\-go\-nal.

By employing our formalism we subsequently explore general properties of FTIs with DCS. {\color{black}Specifically}, we compare the topological properties arising from a DCS with and without a static analog. To clarify this, we point out that in Floquet systems it is possible to preserve the chiral symmetry of the static system by even considering a driving perturbation which commutes with the static Hamiltonian. Such a scenario is unique to driven systems, since a static Hamiltonian is instead required to strictly anticommute with the chiral symmetry operator. From our analysis, we find that DCS with and without static analogs impact the topological properties in a ra\-di\-cal\-ly different fashion, with the latter generally ten\-ding to leave the topological properties of the static system unaffected. Hence, a DCS with no static analog is ideal for topological $\pi$-mode engineering. In fact, we find that the topological properties in this case are mainly determined by the momentum dependence of the driving.

We confirm the above general conclusions obtained from the QEO approach for a concrete 1D chiral p-wave superconductor system where both types of MZMs and MPMs are expected to emerge in the low-frequency dri\-ving regime. This analysis also gives us the opportunity to conduct a comparative ana\-ly\-sis between the here-proposed QEO methodology and the TEO approach, and illustrate the advantages of the QEO method for harmonic drives. Lastly, we establish the connection between the topological invariants obtained using the QEO method and the ones obtained from the effective Floquet Hamiltonian defined in the first Floquet zone (FZ)~\cite{kitagawa2010b,Asboth2012}.

Aside from the case of a BDI system, we further investigate the corresponding consequences of such DCS possibilities for a static 1D system in class CI. While class CI systems are topologically trivial in 1D~\cite{Ryu2010}, adding the driving allows for symmetry breaking transitions, see for instance Ref.~\onlinecite{bandyopadhyay2019}, which can enable the Floquet system to harbor topological edge modes. Indeed, here we demonstrate the case of a symmetry-class conversion CI$\rightarrow$AIII which gives rise to topological $\pi$-modes. Moreover, we find that topological zero modes do not appear, at least in the situations examined here, which is con\-si\-stent with the expectations of our earlier general QEO analysis.

The remainder is organized as follows. In Sec.~\ref{sec:QEOformalism} we review and further expand on the framework of the QEO formalism, while we also detail the notation and conventions to be adopted throughout. Next, in Sec.~\ref{sec:DCS}, we discuss the emergence of DCS and obtain a general formula that leads to the Floquet topological invariants. Using the latter, in Sec.~\ref{sec:Heuristic} we carry out a study of the general properties of driven systems with chiral symmetry, by considering a particular limit, i.e., we focus on two band models and drives with only the principal harmonic of the time-periodic potential. In  Sec.~\ref{sec:NewScenarios} we discuss the effects of the additional presence of antiunitary symmetries in the QEO, and the possibility of symmetry class transitions upon driving. Motivated by these general conclusions of ours, in Secs.~\ref{sec:BDI} and~\ref{sec:CI}, we go on and employ the earlier developed framework to study the effects of driving on the topological properties of a representative static Hamiltonian in class BDI and CI, respectively. In the latter case, we also highlight the symmetry class conversion CI$\rightarrow$AIII and its consequence on the emergence of topological $\pi$-modes.  Section~\ref{sec:Conclusions}, summarizes our findings and concludes our work. Finally, a com\-pa\-ri\-son between the QEO and the TEO methods is discussed in Appendices~\ref{sec:TEOvsQEO} and~\ref{sec:BandInvariants}.

\section{Quasi-Energy Operator Formalism}\label{sec:QEOformalism}

We now proceed by introducing the framework that we employ in order to obtain the Floquet topological inva\-riants $\nu_{0,\pi}\in\mathbb{Z}$ at quasi-energies $\varepsilon=\big\{0,\pi/T\big\}$. Note that throughout this work we strictly consider translationally-invariant driven sy\-stems with chiral symmetry. For simplicity, we focus on 1D sy\-stems, since extensions to higher odd spatial dimensions are straightforward~\cite{Ryu2010}. The starting point of our analysis is the periodically time-dependent Hamiltonian operator:
\begin{align}
H(t,k)=H_{\rm stat}(k)+V(t,k),
\end{align}

\noi which consists of the static part $H_{\rm stat}(k)$ and the periodic driving $V(t+T,k)=V(t,k)$. Here, $k\in[-\pi,\pi)$ defines the quasi-momentum of the band structure of the cry\-stalline system, whose lattice constant is set to unity.

The static part of the Hamiltonian is assumed to feature a chiral symmetry effected by the operator $\Gamma=\Gamma^\dag$, i.e., we have the constraint $\Gamma^\dag H_{\rm stat}(k)\Gamma=-H_{\rm stat}(k)$. In the remainder, we choose a basis which renders the static Hamiltonian block off-diagonal according to:
\begin{align}
H_{\rm stat}(k)=\begin{pmatrix}
0_h&h(k)\\
h^\dag(k)&0_h
\end{pmatrix}\equiv\lambda_+\otimes h(k)+\lambda_-\otimes h^\dag(k)
\end{align}

\noi where we make use of the Pauli matrices $\lambda_{1,2,3}$, the respective unit matrix $\lambda_0$, and the raising/lowering off-diagonal operators $\lambda_\pm=(\lambda_1\pm i\lambda_2)/2$. In the same basis, the chiral symmetry operator takes the block diagonal form $\Gamma=\lambda_3\otimes\mathds{1}_h$. Here, with $0_h$ ($\mathds{1}_h$) we denote the null (identity) operator which is defined in the same space as $h(k)$. Lastly, $\otimes$ denotes the Kronecker product symbol.

After adding the driving, the definition of chiral symmetry needs to be generalized in the following sense~\cite{Asboth2014}: \begin{align}
\Gamma^\dag H(t,k)\Gamma=-H(-t,k)\,.
\label{eq:GenCS}
\end{align}

\noi Hence, for $\Gamma$ to generate a chiral symmetry also in the presence of the drive, each term compri\-sing the dri\-ving Hamiltonian is required to satisfy two combined relations. These pairs of relations either take the form $\Gamma^\dag V(t,k)\Gamma=-V(t,k)$ and $V(-t,k)=+V(t,k)$, or, $\Gamma^\dag V(t,k)\Gamma=+V(t,k)$ and $V(-t,k)=-V(t,k)$. The former case is the extension of the static chiral symmetry condition, while the latter one has no static analog.

In the remainder, we explore qualitative differences in the topological properties of the Floquet insulator resul\-ting from the two distinct DCS scenarios. Before pro\-cee\-ding, we remark that it is convenient to decompose the dri\-ving term according to:
\begin{align}V(t,k)=\sum_{\mu=0,1,2,3}\lambda_\mu\otimes V^{(\mu)}(t,k)\,,\label{eq:DrivingDecomp}
\end{align}

\noi where the driving terms appearing in the above decomposition satisfy $V^{(0),(3)}(-t,k)=-V^{(0),(3)}(t,k)$ and $V^{(1),(2)}(-t,k)=V^{(1),(2)}(t,k)$ so that Eq.~\eqref{eq:GenCS} holds.

Since throughout this work we restrict to harmonic driving terms, it is more suitable to employ the QEO approach. For this purpose, we now make use of the Floquet theorem and label the eigestates $\big|\psi(t,k)\big>$ of $H(t,k)$ with the quasi-energy dispersions $\varepsilon_s(k)$, so that $\big|\psi_s(t,k)\big>=e^{-i\varepsilon_s(k)t}\big|u_s(t,k)\big>$, where we introduced the time-periodic functions $\big|u_s(t+T,k)\big>=\big|u_s(t,k)\big>$. Each $\big|u_s(t,k)\big>$ sa\-ti\-sfies the Floquet quasi-energy equation~\cite{eckardt2015}:
\begin{align}
\big[H(t,k)-i\partial_t\mathds{1}_H\big]\big|u_s(t,k)\big>=\varepsilon_s(k)\big|u_s(t,k)\big>\,.
\label{eq:QEO}
\end{align}

\noi Similar to previous definitions, the identity operator $\mathds{1}_H$ lives in the Hilbert space $\mathbb{H}$ spanned by the matrix $H_{\rm stat}$.

The frequency space quasi-energy formalism is obtained after making use of the Fourier expansion $\big|u_s(t,k)\big>=\sum_ne^{in\omega t}\big|u_{s;n}(k)\big>$, where $\omega=2\pi/T$ is the dri\-ving frequency. This leads to the eigenvalue problem $\sum_mH_{n-m}(k)\big|u_{s;m}(k)\big>=\varepsilon_s(k)\big|u_{s;n}(k)\big>$, where we introduced the ope\-ra\-tor matrix elements:
\begin{align}
H_{n-m}(k)=\big[n\omega\mathds{1}_H+H_{\rm stat}(k)\big]\delta_{n,m}+V_{n-m}(k),
\label{eq:Hmatrixelements}
\end{align}

\noi along with the Fourier components in frequency space:
\begin{align}
V_n(k)=\frac{1}{T}\int_0^Tdt\,e^{-in\omega t}V(t,k),
\end{align}

\noi where $n\in\mathbb{Z}$. Following the decomposition scheme shown in Eq.~\eqref{eq:DrivingDecomp}, we also have $V_n(k)=\sum_{\mu=0,1,2,3}\lambda_\mu\otimes V_n^{(\mu)}(k)$, where $V_n^{(\mu)}(k)=\frac{1}{T}\int_0^Tdt\,e^{-in\omega t}V^{(\mu)}(t,k)$.

We are now in a position to introduce the QEO $\hat{\cal H}_s(k)$ at a given quasi-energy $\varepsilon=s/T$ with $s=\{0,\pi\}$. Each QEO is defined in the extended Floquet-Hilbert space $\mathbb{S}\equiv\mathbb{F}\bigotimes\mathbb{H}$, i.e., the so-called Sambe space~\cite{Sambe}. Here, $\mathbb{F}$ denotes the space spanned by all square-integrable $T$-periodic time-dependent functions~\cite{eckardt2015}. In general, with $\hat{A}$ we denote a matrix named $A$ which lives in $\mathbb{S}$ space. Here, the QEOs are measured relative to the respective quasi-energies. Hence, their matrix elements in $\mathbb{F}$ space take the form:
\begin{align}
{\cal H}_{s;n,m}(k)=\Omega_{s;n,m}+H_{\rm stat}(k)\delta_{n,m}+V_{n-m}(k),
\label{eq:QEO}
\end{align}

\noi where we set:
\begin{align}
\Omega_{s;n,m}\equiv\Omega_{s;n-m}=\left(n-\frac{s}{2\pi}\right)\omega\,\delta_{n,m}\mathds{1}_H\,.
\end{align}

Each QEO is a hermitian matrix in $\mathbb{S}$. This is because Floquet complex conjugation (FCC), that we define as $V_n^\star(k)\equiv V_{-n}(k)$, is equivalent to hermitian conjugation $V_n^\star(k)=V_n^\dag(k)$. Note also that the shift by $s\omega/2\pi$ introduced in each $\hat{\cal H}_s(k)$ sets the respective quasi-energy to zero. This is convenient, since it implies that to\-po\-lo\-gi\-cal gap closings at each quasi-energy occur when $\hat{\cal H}_s(k)$ is zero. For the $k$ points that this happens, the related to\-po\-lo\-gi\-cal invariants are ill defined and mark the topological phase transition boundaries in parameter space.

\section{Dynamical chiral symmetry}\label{sec:DCS}

In order to obtain the Floquet topological invariants, it is required to determine the form of operator $\hat{\Pi}_s=\hat{\Pi}_s^\dag$ which generates the DCS for the respective quasi-energy and implies the relation $\hat{\Pi}_s^\dag\hat{\cal H}_s(k)\hat{\Pi}_s=-\hat{\cal H}_s(k)$. To achieve this, it is first crucial to identify operators which anticommute with the matrices $\hat{\Omega}_s$, which have a nontrivial structure solely in $\mathbb{F}$ space. {\color{black}To gain further insight, we show below the matrix structures of $\hat{\Omega}_{0,\pi}$:
\bea
\frac{\hat{\Omega}_0^{\rm diag}}{\omega}&=&
\begin{pmatrix}
+\check{N}&\bm{0}_N&\check{0}_N\\
\bm{0}_N^\intercal&0&\bm{0}_N^\intercal\\
\check{0}_N&\bm{0}_N&-\check{N}\\
\end{pmatrix}\otimes\mathds{1}_H\quad{\rm and}\quad\\
\frac{\hat{\Omega}_\pi^{\rm diag}}{\omega}&=&
\begin{pmatrix}
+\left(\check{N}-\nicefrac{1}{2}\check{\mathds{1}}_N\right)
&\check{0}_N\\
\check{0}_N&-\left(\check{N}-\nicefrac{1}{2}\check{\mathds{1}}_N\right)
\end{pmatrix}\otimes\mathds{1}_H,\qquad\label{eq:OmegapiOriginal}
\eea

\noi which are here compactly expressed as block diagonal matrices. To achieve this, we employed suitable zero and $\pi$ quasi-energy bases, for which, the quantum numbers $n$ and $m$ appearing in Eq.~\eqref{eq:Hmatrixelements} take values defined by the sets $\big\{\check{N},0,-\check{N}\big\}$ and $\big\{\check{N},-\check{N}\big\}$, respectively. In the above, we introduced the operator $\check{N}$, which in its eigenbasis is represented by the infinite square dia\-go\-nal matrix ${\rm diag}\{\infty,\ldots,3,2,1\}$. Further, $\bm{0}_N^\intercal$ yields the row vector $\{0,\ldots,0,0,0\}$ with $^\intercal$ denoting matrix transposition in the space defined by $\check{N}$. In addition, $\check{0}_N$ and $\check{\mathds{1}}_N$ define square null and identity matrices in the eigenbasis of $\check{N}$. Note that the \textit{sequence} of values for $n$ and $m$ obtained from the above bases differs compared to the usual one, in which $n,m=\{+\infty,\ldots,3,2,1,0,-1,-2,-3,\ldots,-\infty\}$.

In order to proceed with the identification of the ope\-ra\-tors $\hat{\Pi}_s$ establishing the DCS for the QEOs, it is more convenient to transfer to a new basis, in which the matrices $\hat{\Omega}_s$ are represented by block off-diagonal matrices and, at the same time, $\hat{\Pi}_s$ become diagonal. In particular, we transfer to bases which lead to the expressions:
\bea
\frac{\hat{\Omega}_0}{\omega}&=&
\begin{pmatrix}
\check{0}_N&\bm{0}_N&\check{N}\\
\bm{0}_N^\intercal&0&\bm{0}_N^\intercal\\
\check{N}&\bm{0}_N&\check{0}_N\\
\end{pmatrix}\otimes\mathds{1}_H\quad{\rm and}\quad\\
\frac{\hat{\Omega}_\pi}{\omega}&=&
\begin{pmatrix}
\check{0}_N&\check{N}-\nicefrac{1}{2}\check{\mathds{1}}_N
\\
\check{N}-\nicefrac{1}{2}\check{\mathds{1}}_N&\check{0}_N
\end{pmatrix}\otimes\mathds{1}_H.\label{eq:Omegapi}
\eea

The above representation of $\hat{\Omega}_0$ has been given in terms of the following basis:
\begin{align}
\left\{\frac{\big|\check{N}_0,k\big>+\big|-\check{N}_0,k\big>}{\sqrt{2}},\big|0,k\big>,
\frac{\big|\check{N}_0,k\big>-\big|-\check{N}_0,k\big>}{\sqrt{2}}\right\},\label{eq:BasisZero}
\end{align}

\noi which is expressed in a condensed manner with the help of a new number operator labeled $\check{N}_0$. Similarly, $\hat{\Omega}_\pi$ is defined in the basis:
\begin{align}
\left\{\frac{\big|\check{N}_\pi,k\big>+\big|-\check{N}_\pi,k\big>}{\sqrt{2}},
\frac{\big|\check{N}_\pi,k\big>-\big|-\check{N}_\pi,k\big>}{\sqrt{2}}\right\},\label{eq:BasisPi}
\end{align}

\noi where we introduced a respective number ope\-ra\-tor $\check{N}_\pi$. These two new operators are given by the expressions:
\begin{align}
\check{N}_s=\check{N}-\frac{s}{2\pi}\check{\mathds{1}}_N
\end{align}

\noi with $s=\{0,\pi\}$. Each operator yields the number of ener\-gy quanta in units of $\omega$ that the drive exchanges with the FTI for a given quasi-energy sector.

At this point, it is crucial to note that the bases in Eqs.~\eqref{eq:BasisZero} and~\eqref{eq:BasisPi} are now defined in terms of a set of eigenstates instead of being labeled by the values of the number operator $\check{N}$. Specifically, in the above, we made use of the shorthand formulas:
\begin{align}
\big|\check{N}_s,k\big>=
\big|u_{s=0;n-s/2\pi}(k)\big>.
\end{align}

\noi Employing the above basis notation allows us to transparently illustrate that the new basis is obtain by making linear combinations of states which possess opposite va\-lues for the number operator of each quasi-energy sector, i.e., linear combinations of the form $\big|\check{N}_s,k\big>\pm\big|-\check{N}_s,k\big>$.}

By means of the above basis choices, we infer that the components of $\hat{\Omega}_{0,\pi}$ which are defined in $\mathbb{F}$ space correspondingly anticommute with the matrices $\tilde{\cal F}_{0,\pi}$ which solely act in $\mathbb{F}$ space and read as follows:
\begin{align}
\tilde{\cal F}_0=
\begin{pmatrix}
\check{\mathds{1}}_N&\bm{0}_N&\check{0}_N\\
\bm{0}_N^\intercal&1&\bm{0}_N^\intercal\\
\check{0}_N&\bm{0}_N&-\check{\mathds{1}}_N\\
\end{pmatrix}\phd{\rm and}\phd\,
\tilde{\cal F}_\pi=
\begin{pmatrix}
\check{\mathds{1}}_N&\check{0}_N\\
\check{0}_N&-\check{\mathds{1}}_N
\end{pmatrix}.
\end{align}

\noi From the above expressions, we find that each $\tilde{\cal F}_s$ has an effect which is equivalent to inverting the respective number ope\-ra\-tors, i.e., $\check{N}_s\mapsto-\check{N}_s$. Therefore, the ope\-ra\-tion effected by $\tilde{\cal F}_s$ is equivalent to FCC and, in the real time domain, to the inversion $t\mapsto-t$.

The above considerations allow us to directly identify the DCS operators with $\hat{\Pi}_s=\tilde{\cal F}_s\otimes\Gamma$, since these clearly anticommute with $\hat{\Omega}_s$ and $\mathds{1}_F\otimes H_{\rm stat}(k)$, where $\mathds{1}_F$ corresponds to the identity matrix in $\mathbb{F}$ space. The $\hat{\Pi}_s$ ope\-ra\-tors also anticommute with the driving perturbation matrices in $\mathbb{S}$, that we here denote $\hat{\cal V}_s(k)$. The latter property results from the transformation assumed here for $V(t,k)$ under the combined action of FCC and the static chiral symmetry operator $\Gamma$.

In the Floquet bases chosen {\color{black} above and kept throughout the remainder of} this work, the driving terms defined in $\mathbb{S}$ space take the general form:
\bea
\hat{\cal V}_0(k)&=&
\begin{pmatrix}
\check{U}_+(k)&\bm{v}(k)&i\,\check{\cal U}(k)\\
\bm{v}^\intercal(k)&0&-i\,\bm{u}^\intercal(k)\\
-i\,\check{\cal U}^\intercal(k)&i\,\bm{u}(k)&\check{U}_-(k)\\
\end{pmatrix},\quad\\
\hat{\cal V}_\pi(k)&=&
\begin{pmatrix}
\check{U}_+(k)&i\,\check{\cal U}(k)\\
-i\,\check{\cal U}^\intercal(k)&\check{U}_-(k)\\
\end{pmatrix}.\label{eq:Vpi}
\eea

\noi In the above, we introduced the row vector operators $\bm{v}^\intercal(k)$ and $\bm{u}^\intercal(k)$, which have entries:
\begin{align}
v_a(k)=\frac{V_a(k)+V_a^\star(k)}{\sqrt{2}}\phd{\rm and}\phd
u_a(k)=\frac{V_a(k)-V_a^\star(k)}{\sqrt{2}i},
\end{align}

\noi where depending on the quasi-energy sector $s=\{0,\pi\}$, the index $a$ takes values $a=\big\{\check{N}_s\big\}\equiv\big\{\infty,\ldots,3-s/2\pi,2-s/2\pi,1-s/2\pi\big\}$. Moreover, we defined $\check{U}_\pm(k)$ and $\check{\cal U}(k)$, which constitute square-matrix operators with matrix ele\-ments:
\bea
U_{\pm;a,b}(k)&=&\frac{V_{a-b}(k)+V_{a-b}^\star(k)}{2}\pm\frac{V_{a+b}(k)+V_{a+b}^\star(k)}{2}\,,\no\\\\
{\cal U}_{a,b}(k)&=&\frac{V_{a-b}(k)-V_{a-b}^\star(k)-\big[V_{a+b}(k)-V_{a+b}^\star(k)\big]}{2i}\,.\no\\
\eea

Note that the indices $a,b$ take values in the set $\big\{\check{N}_s\big\}$ depending on the quasi-energy sector $s=\{0,\pi\}$. We can further decompose all the above quantities in the same fashion as it was done in Eq.~\eqref{eq:DrivingDecomp}, i.e., we can write them as $\bm{v}(k)=\sum_{\mu=0,1,2,3}\lambda_\mu\otimes\bm{v}^{(\mu)}(k)$, $\bm{u}(k)=\sum_{\mu=0,1,2,3}\lambda_\mu\otimes\bm{u}^{(\mu)}(k)$, $\check{U}_\pm(k)=\sum_{\mu=0,1,2,3}\lambda_\mu\otimes\check{U}_\pm^{(\mu)}(k)$, and $\check{\cal U}(k)=\sum_{\mu=0,1,2,3}\lambda_\mu\otimes\check{\cal U}^{(\mu)}(k)$. Due to the assumed ``locked" behavior of the driving terms under FCC and the chiral symmetry effected by $\Gamma$, a number of the quantities $\bm{v}^{(\mu)}$, $\bm{u}^{(\mu)}$, $\check{U}_\pm^{(\mu)}$, and $\check{\cal U}^{(\mu)}$ will even\-tual\-ly va\-nish, thus leading to simplified expressions for the driving matrices of a given matrix $\lambda_{0,1,2,3}$.

At this stage, it is important to remark that by virtue of the Floquet basis chosen earlier, $\hat{\Pi}_s$ are dia\-gonal matrices. {\color{black}However, their diagonal elements are not yet ordered into two blocks with $\pm1$ values}, which is actual\-ly desired for obtaining the topological invariants. To achieve this, it is required to employ suitable unitary transformations effected by $\hat{\cal O}_s(k)$, which bring the DCS operators $\hat{\Pi}_s$ in the desired block dia\-gonal form mentioned above. At the same time, each unitary transformation block off-diagonalizes the respective QEO according to $\underline{\hat{\cal H}}_s(k)=\hat{\cal O}_s^\dag(k)\hat{\cal H}_s(k)\hat{\cal O}_s(k)$. While there is no unique choice for the unitary transformations which allow us to obtain the block off-diagonal forms shown {\color{black}later on}, we here consider concrete transformations which lead to $\hat{\Pi}_0={\rm diag}\big\{\check{\mathds{1}}_N,\check{\mathds{1}}_N,1,-1,-\check{\mathds{1}}_N,-\check{\mathds{1}}_N\big\}\otimes\mathds{1}_h$ and $\hat{\Pi}_\pi={\rm diag}\big\{\check{\mathds{1}}_N,\check{\mathds{1}}_N,-\check{\mathds{1}}_N,-\check{\mathds{1}}_N\big\}\otimes\mathds{1}_h$, and read as:
\bea
\hat{\cal O}_0&=&
\begin{pmatrix}
\check{\mathds{1}}_N&\check{0}_N&\bm{0}_N&\bm{0}_N&\check{0}_N&\check{0}_N\\
\check{0}_N&\check{0}_N&\bm{0}_N&\bm{0}_N&\check{0}_N&\check{\mathds{1}}_N\\
\bm{0}^\intercal_N&\bm{0}^\intercal_N&1&0&\bm{0}^\intercal_N&\bm{0}^\intercal_N\\
\bm{0}^\intercal_N&\bm{0}^\intercal_N&0&1&\bm{0}^\intercal_N&\bm{0}^\intercal_N\\
\check{0}_N&\check{0}_N&\bm{0}_N&\bm{0}_N&\check{\mathds{1}}_N&\check{0}_N\\
\check{0}_N&\check{\mathds{1}}_N&\bm{0}_N&\bm{0}_N&\check{0}_N&\check{0}_N\\
\end{pmatrix}\otimes\mathds{1}_h\,,\\
\no\\
\hat{\cal O}_\pi&=&
\begin{pmatrix}
\check{\mathds{1}}_N&\check{0}_N&\check{0}_N&\check{0}_N\\
\check{0}_N&\check{0}_N&\check{0}_N&\check{\mathds{1}}_N\\
\check{0}_N&\check{0}_N&\check{\mathds{1}}_N&\check{0}_N\\
\check{0}_N&\check{\mathds{1}}_N&\check{0}_N&\check{0}_N\\
\end{pmatrix}\otimes\mathds{1}_h\,.
\eea

\noi The above transformation result in the following upper off-diagonal blocks for the QEOs:
\bea
&&\hat{A}_0=
\begin{pmatrix}
\bm{0}_N\otimes\mathds{1}_h&\omega{\color{black}\check{N}_0}\otimes\mathds{1}_h&\check{\mathds{1}}_N\otimes h\\
\bm{0}_N\otimes\mathds{1}_h&\check{\mathds{1}}_N\otimes h^\dag&\omega{\color{black}\check{N}_0}\otimes\mathds{1}_h\\
h&\bm{0}^\intercal_N\otimes\mathds{1}_h&\bm{0}^\intercal_N\otimes\mathds{1}_h
\end{pmatrix}+\no\\
&&\begin{pmatrix}
\bm{v}^{(1)}-i\bm{v}^{(2)}&i\big[\check{\cal U}^{(0)}+\check{\cal U}^{(3)}\big]&\check{U}_+^{(1)}-i\,\check{U}_+^{(2)}\\
i\big[\bm{u}^{(0)}-\bm{u}^{(3)}\big]&\check{U}_-^{(1)}+i\,\check{U}_-^{(2)}&-i\big[\check{\cal U}^{(0)}-\check{\cal U}^{(3)}\big]^\intercal\\
0&-i\big[\bm{u}^{(0)}+\bm{u}^{(3)}\big]^\intercal&\big[\bm{v}^{(1)}-i\bm{v}^{(2)}\big]^\intercal
\end{pmatrix},\no\\\label{eq:Azero}\\
&&\hat{A}_\pi=
\begin{pmatrix}
\omega{\color{black}\check{N}_\pi}\otimes\mathds{1}_h&\check{\mathds{1}}_N\otimes h\\
\check{\mathds{1}}_N\otimes h^\dag&\omega{\color{black}\check{N}_\pi}\otimes \mathds{1}_h\\
\end{pmatrix}\no\\
&&\phd\quad+\begin{pmatrix}
i\big[\check{\cal U}^{(0)}+\check{\cal U}^{(3)}\big]&\check{U}_+^{(1)}-i\,\check{U}_+^{(2)}\\
\check{U}_-^{(1)}+i\,\check{U}_-^{(2)}&-i\big[\check{\cal U}^{(0)}-\check{\cal U}^{(3)}\big]^\intercal
\end{pmatrix},\label{eq:Api}
\eea

\noi where in the above we suppressed the quasi-momentum argument to simplify the notation.

After effecting these unitary transformations we end up with the fol\-lowing matrix structures:
\begin{align}
\underline{\hat{\cal H}}_s(k)=
\begin{pmatrix}
\hat{0}&\hat{A}_s(k)\\
\hat{A}_s^\dag(k)&\hat{0}\end{pmatrix}
\phd{\rm and}\phd
\underline{\hat{\Pi}}_s=
\begin{pmatrix}
\hat{\mathds{1}}&\hat{0}\\
\hat{0}&-\hat{\mathds{1}}\end{pmatrix}\,.\label{eq:BlockOffDiag}
\end{align}

With $\hat{A}_{0,\pi}(k)$ at hand, we identify the Floquet topological invariants in 1D as the winding numbers:
\begin{align}
\nu_s=\frac{i}{2\pi}\int_{-\pi}^{+\pi}dk\ph{\rm tr}\left[\hat{A}_s^{-1}(k)\frac{d\hat{A}_s(k)}{dk}\right],
\end{align}

\noi where ``{\rm tr}" denotes matrix trace. After the analysis of Refs.~\cite{Ryu2010}, extensions to higher odd dimensions are straightforward. Note that in 1D the above expression can be simplified by making use of the identity ${\rm tr}\ln\equiv\ln\det$ and introducing the determinant:
\[
\det[\hat{A}_s(k)]=\big|\det[\hat{A}_s(k)]\big|e^{-i\varphi_s(k)}\,.\]

\noi {\color{black} Rewriting $\nu_s$ in terms of the determinant $\det[\hat{A}_s(k)]$ also illustrates in a transparent fashion that the winding number is well-defined as long as the modulus of the determinant is nonzero for all $k$. Since we here focus on insulators, this requirement is always satisfied.

Aside from the above, it is worthy to note that the steps discussed earlier} immediately lead to the compact expression $\nu_s=\int_0^{2\pi}\, d\varphi_s/2\pi$. Therefore, the arising topological phases can be clasified by obser\-ving how the phase $\varphi_s(k)$ evolves as $k$ sweeps through the Brillouin zone~\cite{Ryu2010}.

\section{Heuristic Study of Topological Properties Upon Driving}\label{sec:Heuristic}

From the expressions for $\hat{A}_{0,\pi}$ obtained above, we observe that driving terms with the same transformation properties under the action of FCC bunch together. Hence, those with a different transformation appear in different ``blocks" in the $\hat{A}_{0,\pi}(k)$ matrices. Due to this tendency, we infer that the main effect of terms which are odd (even) under FCC, is to act in the number $\check{N}$ (Hamiltonian-block $h$) space. Hence, already from this, we anticipate that the two classes of driving terms also lead to qualitative different topological effects.

This can be shown heuristically by considering a particular limit, in which, only two bands of $H_{\rm stat}(k)$ become relevant for the topological properties of both the static and driven system. {\color{black} This limit is of interest when one investigates a topological phase transition taking place via the gap closings and re-openings in the vicinity of a specific $k$.} Under, this condition, the $h(k)$ ope\-ra\-tor simply becomes a complex function. By further trun\-ca\-ting the eigenspace spanned by $\check{N}$ to the minimal case of $\check{N}=1$, and considering that only the component $\hat{V}_1(k)$ of the driving contributes, we are in a position to obtain a list of analytical conclusions {\color{black} that we present in detail in the next subsections. Before that, however, it is important to note that the above truncation is equivalent to employing the so-called ro\-ta\-ting wave approximation, see for instance Ref.~\onlinecite{lindner2011}. Such an approximation needs to be treated with caution, and it is expected to work well when the frequency $\omega$ is larger than the band gap of the static system and at the same time smaller than its bandwidth. In addition, to further guarantee the validity of this approach, it is important that multiples of $\omega$ exceed the bandwidth, while it is also crucial that the respective driving perturbation $\hat{V}_1(k)$ hybridizes states predominantly belon\-ging to the two bands of interest.}

\subsection{Even-under-FCC driving}

As mentioned above, we distinguish two cases depen\-ding on the behavior of the driving term under FCC. We first examine the case of DCS which has a static analog, and is a result of driving terms which are invariant under FCC.

\subsubsection{Zero Quasi-Energy Topology}

Under the approximations discussed at the beginning of this section, we find that the only driving terms contributing are $\bm{v}^{(1),(2)}(k)$. By restricting to $\check{N}=1$, we end up with the complex function $\bm{v}^{(1)}(k)-i\bm{v}^{(2)}(k)\mapsto v^{(1)}(k)-iv^{(2)}(k)\equiv|v(k)|e^{-i\theta(k)}$. We further set $h(k)=|h(k)|e^{-i\varphi_{\rm stat}(k)}$, where $\varphi_{\rm stat}(k)$ is the phase factor whose winding number $\nu_{\rm stat}=\int_0^{2\pi}d\varphi_{\rm stat}/2\pi$ determines the static topological invariant $\nu_{\rm stat}$. From Eq.~\eqref{eq:Azero} we find the following result for the determinant of $\hat{A}_0(k)$:
\bea
\det\big[\hat{A}_0\big]&=&
|h|e^{-i\varphi_{\rm stat}}\Big\{\omega^2-|h|^2+|v|^2\cos\big[2(\varphi_{\rm stat}-\theta)\big]\no\\
&+&i|v|^2\sin\big[2(\varphi_{\rm stat}-\theta)\big]\Big\}\,,
\eea

\noi where in the above we suppressed the argument of the quasi-momentum to simplify the notation. Such simplifications will take place throughout this work from now on without any further mentioning.

From the resulting expression for the determinant, we observe that the operator $h$ of the static Hamiltonian fully factors out. One hand, this establishes the smooth limit to the static case which is obtained by consi\-de\-ring $|v(k)|\rightarrow0$ for all $k$. Furthermore, it implies that the Floquet topological invariant at zero quasi-energy is given by the sum of the static invariant $\nu_{\rm stat}$ and a new contribution, which stems from the winding of the argument of the complex number in brackets of the above expression. A nonzero winding for the latter term can arise when the relations $|v|^2\sin\big[2(\varphi_{\rm stat}-\theta)\big]=0$ and $\omega^2+|v|^2\cos\big[2(\varphi_{\rm stat}-\theta)\big]=|h|^2$ can become simultaneously sa\-ti\-sfied for a given $k$. {\color{black}Satisfying the two conditions marks the appearance of additional band tou\-chings which, in turn, can mediate topological phase transitions which are inaccesible in the static case. Previously not possible topological phases and different values for the winding numbers can emerge when a gap opens at these additional band touching points and the two conditions discussed above are no longer met.} There exist two possibi\-li\-ties: (i) $|v(k)|=0$, in which case $\omega^2=|h(k)|^2$ needs to be fulfilled, or, (ii) $\varphi_{\rm stat}(k)-\theta(k)=\{0,\pm\pi/2,\pi\}$ where now the constraint $\omega^2\pm|v(k)|^2=|h(k)|^2$ has to be met. For a frequency which is sufficiently larger that the static's sy\-stems energy gap, it is scenario (ii) that is the most pro\-ba\-ble to take place, since the driving term effectively reduces the frequency. {\color{black}In either case, the above imply that} a threshold strength for the driving is required in order to modify the topological properties of the system.

\subsubsection{$\pi$ Quasi-Energy Topology}\label{sec:piEvenFCCanalytics}

We now proceed with examining the topological pro\-per\-ties that are induced at $\pi$ quasi-energy, due to driving terms which are even under FCC.

The restriction to $\check{N}=1$ implies that, in turn, the corresponding number operator becomes $\check{N}_\pi=\nicefrac{1}{2}$. This further yields that the contributing driving terms $\check{U}_\pm^{(1),(2)}(k)$ can be replaced by the real functions $U_\pm^{(1),(2)}$ which sa\-ti\-sfy $U_\pm^{(1),(2)}=-U_\mp^{(1),(2)}$. Therefore, in ana\-lo\-gy to the previous paragraph, in the following we set $U_+^{(1)}(k)-iU_+^{(2)}(k)\equiv|U(k)|e^{-i\theta(k)}$. From Eq.~\eqref{eq:Api} we now obtain:
\begin{align}
\det\big[\hat{A}_\pi\big]=(\omega/2)^2-|h|^2+|U|^2-2i|h||U|\sin\big(\varphi_{\rm stat}-\theta\big).
\end{align}

\noi In the above, we first observe that in the static limit the determinant becomes real, which is consistent with the expected absence of topologically nontrivial phases at $\pi$ quasi-energy. Assuming that the static system is an insulator with $|h|\neq0$, we conclude that to\-po\-lo\-gi\-cal phase transitions occur in the presence of driving when (i) $|U(k)|=0$ and $\omega/2=|h(k)|$ are simultaneously satisfied, or, (ii) when $\varphi_{\rm stat}(k)=\theta(k)+\big\{0,\pi\big\}$ and $(\omega/2)^2+|U(k)|^2=|h(k)|^2$ are fulfilled in tandem. In either case, gaps can open at bands crossing $\pi$ quasi-energy even for infinitesinally weak drivings, hence topologically nontrivial phases become immediately accessible upon switching on the driving perturbation.

\subsection{Odd-under-FCC driving}

We now proceed with the second class of driving terms and examine their effects on the topological properties at the two quasi-energies of interest.

\subsubsection{Zero Quasi-Energy Topology}

In this case, only $\bm{u}^{(0),(3)}(k)$ survive and from vectors become real functions that we label as $u^{(0)}(k)=u_{\mathds{1}}(k)$ and $u^{(3)}(k)=u_{\Gamma}(k)$. Note that the indices have been chosen so to reflect the matrix structure of the driving terms. The determinant of $\hat{A}_0(k)$ now becomes:
\begin{align}
\det\big[\hat{A}_0\big]=
|h|e^{-i\varphi_{\rm stat}}\Big(\omega^2-|h|^2+u_{\mathds{1}}^2-u_{\Gamma}^2\Big)\,.
\label{eq:TrendsZeroOdd}
\end{align}

\noi From the above, we immediately find that the topological properties at zero quasi-energy do not differ compared to the static ones.

\subsubsection{$\pi$ Quasi-Energy Topology}

We now consider the effects on the other quasi-energy. Here, the matrices $\check{\cal U}^{(0),(3)}(k)$ become relevant. Under the assumption of the exchange of a single quantum of energy with the drive, these become real functions that we denote ${\cal U}_{\mathds{1}}(k)$ and ${\cal U}_{\Gamma}(k)$, respectively. Within these definitions in place, we now find:
\begin{align}
\det\big[\hat{A}_\pi\big]=(\omega/2)^2-|h|^2+{\cal U}_{\mathds{1}}^2-{\cal U}_{\Gamma}^2+i\omega\,{\cal U}_{\Gamma}.
\label{eq:TrendsPiOdd}
\end{align}

\noi Quite remarkably, topological gaps open at quasi-momenta which simultaneously satisfy the relations ${\cal U}_{\Gamma}(k)=0$ and $(\omega/2)^2+\big[{\cal U}_{\mathds{1}}(k)\big]^2=|h(k)|^2$. This implies that a term which is proportional to the identity matrix cannot induce any nontrivial phases, in contrast to a driving term which possesses a matrix structure that coincides with the static chiral symmetry matrix $\Gamma$. In fact, when only ${\cal U}_{\Gamma}(k)$ is nonzero, the driving opens immediately gaps when there exist bands crossing $\pi$ quasi-energy. Whether, a nonzero $\nu_\pi$ is induced will depend on the precise $k$-dependence of $h(k)$ and ${\cal U}_{\Gamma}(k)$.

\section{Topological Scenarios without static analogs}\label{sec:NewScenarios}

From the results of the previous section, we conclude that driving terms which are odd under FCC have little impact on the topological properties at zero quasi-energy, while the topological properties at $\pi$ quasi-energy are strongly decided by the Hamiltonian structure of the driving perturbation. Despite the fact that these conclusions were inferred in a particular limit and under a specific truncation of the QEOs, we expect these trends to also hold in more general situations, at least, as long as the strength of the driving perturbation is weak.

These arguments illustrate that driving terms which are odd under FCC are most prominent for leading to effects without a static counterpart, since these are not constrained to commute with the operator generating the chiral symmetry of the static system. Remarkably, such a loophole opens the door to driving-induced symmetry class transitions and a concomitant rich topological phenomenology. On these grounds, in the remainder of this work we primarily concentrate on the effects of such nonstandard topological scenarios which arise from dri\-ving terms which are odd under FCC.

For this purpose, we return to the QEO for $\pi$ quasi-energy, and analyze its symmetry and topological pro\-per\-ties. It is more convenient to carry out this program in the original representation of the QEO, i.e., defined using Eqs.~\eqref{eq:Hmatrixelements},~\eqref{eq:Omegapi}, and~\eqref{eq:Vpi}. Hence, in case of the odd-under-FCC driving the $\pi$ quasi-energy QEO takes the following compact form:
\bea
\hat{\cal H}_\pi(k)&=&
\mathds{1}_F\otimes H_{\rm stat}(k)+\kappa_1\otimes{\color{black}\big(\omega\check{N}_\pi\otimes\mathds{1}_H\big)}\no\\
&+&\kappa_1\otimes i\,\check{\cal U}_A(k)+\kappa_2\otimes\check{\cal U}_S(k)\,,
\eea

\noi where we compactly expressed the QEO in the basis of Eq.~\eqref{eq:BasisPi} using the Pauli matrices $\kappa_{1,2,3}$ along with the respective identity matrix $\kappa_0$. {\color{black}We remark that the va\-rious Hamiltonian terms possess the same matrix dimensions in spite of the seemingly nonmatching number of Kronecker pro\-duct operations. For instance, note that $\mathds{1}_F\equiv\kappa_0\otimes\check{\mathds{1}}_N$.} In addition, we introduced the symmetric and antisymmetric matrix operators $\check{\cal U}_S(k)$ and $\check{\cal U}_A(k)$ which have matrix elements given by the expressions:
\bea
\check{\cal U}_{S;a,b}(k)&=&\frac{V_{a+b}(k)-V_{a+b}^\star(k)}{2i}\,,\\
\check{\cal U}_{A;a,b}(k)&=&\frac{V_{a-b}(k)-V_{a-b}^\star(k)}{2i}\,,
\eea

\noi where the driving terms strictly commute with the static chiral symmetry operator, i.e., $\big[V_a(k),\Gamma\big]=0$ for all va\-lues of the index $a$. Therefore, in this basis, the DCS operator is identified with $\hat{\Pi}_\pi=\kappa_3\otimes\check{\mathds{1}}_N\otimes\Gamma$.

So far in this work, we have mainly focused on the chiral symmetry of both the static and driven systems. However, the static Hamiltonian may preserve additional antiunitary symmetries which may persist or get violated upon switching on the driving. There exist two types of such additional symmetries~\cite{AZ}: (i) a generalized time-reversal symmetry effected by the operator $\Theta$, so that $\Theta^\dag H_{\rm stat}(k)\Theta=H_{\rm stat}(-k)$, and, (ii) a charge-conjugation symmetry induced by the operator $\Xi$ so that the fol\-lowing relation holds $\Xi^\dag H_{\rm stat}(k)\Xi=-H_{\rm stat}(-k)$. We point out that since a chiral symmetry is always assumed to be preserved for the static system, if at least a time-reversal or a charge-conjugation symmetry is additionally present, then the static system preserves simultaneously both types of antiunitary symmetries with $\Gamma\propto\Theta\Xi$~\cite{Ryu2010}.

The assumption of chiral symmetry sets constraints on the symmetry classes which are possible for the static system. When only chiral symmetry exists, then static and driven Hamiltonians belong to the AIII symmetry class, which admits a $\mathbb{Z}$ topological invariant in odd spatial dimensions~\cite{Ryu2010}. The other possibilities include the fol\-lo\-wing two pairs of symmetry classes: (i) $\big\{$BDI, CI$\big\}$ with $\Theta^2=\mathds{1}_H$ and $\Xi^2=\pm\mathds{1}_H$, along with (ii) $\big\{$DIII, CII$\big\}$ with $\Theta^2=-\mathds{1}_H$ and $\Xi^2=\pm\mathds{1}_H$, respectively. When the driving terms do not preserve any additional symmetry, then driving the system results in a symmetry collapse of the form $\{$AIII, BDI, CI, DIII, CII$\}\rightarrow$AIII.

At this point, it is crucial to remark that while this collapse rule is generally possible, in certain condensed matter sy\-stems such a symmetry-class breaking scheme is not {\color{black}accessible}. This is because topological superconductors in classes BDI and DIII have a built-in charge-conjugation symmetry~\cite{AZ} which, along with the here-assumed DCS, restricts the driving terms in such a fa\-shion so that the symmetry classes of static and driven systems are identical. Therefore, a symmetry class transition to an AIII class in topological superconductors is mainly accessible in classes CI and CII, which lack such additional built-in antiunitary symmetries. Interestingly, such a symmetry conversion induced by the driving can enable a static system residing in a trivial symmetry class for a given  spatial dimension, to a driven system with nontrivial topological properties. We examine situations belonging to both types in the upcoming sections. Spe\-ci\-fi\-cal\-ly, we study two concrete mo\-dels in 1D. One in which the static system belongs to class BDI and one to class CI. For both models, we investigate the resulting topological properties of the system when this is subject to driving terms which are both even and odd under FCC.

Concluding this section, we note that within the QEO formalism the additional antiunitary symmetries of the static system, which are effected by $\Theta$ and $\Xi$, persist also in the driven state when all the odd-under-FCC potential components $V_a(k)$ change sign under $\Theta$. Hence, only when $\Theta^\dag V_a(k)\Theta=-V_a(-k)$. This can be naturally understood by taking into account that the definitions of the time-reversal and charge-conjugation symmetries for a driven systems need to be extended according to:
\bea
\Theta^\dag H(t,k)\Theta&=&+H(-t,-k)\,,\\
\Xi^\dag H(t,k)\Xi&=&-H(t,-k)\,.
\eea

\noi Together with the definition of DCS in Eq.~\eqref{eq:GenCS}, the extensions of the static symmetries to the driven case are chosen in such a way, so that the Floquet operator, that is defined as ${\cal H}(t,k)=H(t,k)-i\partial_t\mathds{1}_H$, is invariant under these when they are present. For further details on the topological classification of FTIs we prompt the reader to refer to the works in Refs.~\onlinecite{kitagawa2010b,Nathan,fulgab2016,Fruchart2016,potter2016,RoyHarper1,RoyHarper2,Yao2017}. Finally, note that symmetries beyond the tenfold classification scheme have been also discussed in FTIs, such as, in Refs.~\onlinecite{Peng2020,Molignini,Jangjan,Na2023}.

\section{Analysis of a representative BDI model in 1D}\label{sec:BDI}

In this section, we investigate the consequences of harmonic drives, which are both even- and odd-under FCC, on the topological properties of an extended Kitaev chain model in BDI symmetry class. This model is described by the static Bogoliubov-de Gennes (BdG) Hamiltonian:
\begin{align}
H_{\rm BDI}(k)=\varepsilon(k)\tau_3+\Delta(k)\tau_2,
\label{eq:BDIHam}
\end{align}

\noi where we made use of the $\tau_{1,2,3}$ Pauli matrices which act in the Nambu space spanned by the electron $\big|e;k,\uparrow\big>$ and hole states $\big|h;-k,\uparrow\big>$. These two states correspond to the eigenstates $\tau_3=\pm1$, respectively. In addition, we introduced the energy dispersion $\varepsilon(k)=\mu-\cos k-\alpha\cos(2k)$ and the p-wave pairing gap $\Delta(k)=\Delta\sin k \big(1-\beta\cos k\big)$. In the above, $\mu$ defines the che\-mi\-cal potential, $\Delta$ is a pai\-ring energy scale that is set as the energy unit in the remainder, while $\alpha$ and $\beta$ describe the second neighbor hopping amplitude
and pairing potential, respectively. Here, the arising harmonics $\sin k$ and $\sin(2k)$ indicate the nearest and next nearest neighbor Cooper pairing terms. Note that a number of variants of longer-ranged Kitaev chain models have been previously studied for both static~\cite{Mercaldo2016,Patrick,Mercaldo2018,alecce2017,kotetes2022} and driven~\cite{li2017,li2018} systems.

The static energy di\-spersions are of the form $\pm E(k)$ with $E(k)=\sqrt{\varepsilon^2(k)+\Delta^2(k)}$, and exhibit particle-hole symmetry due to a chiral symmetry effected by $\tau_1$. As a consequence, the Hamiltonian can be brought to the block off-diagonal form $H_{\rm stat}(k)=R^\dag H_{\rm BDI}(k)R$ with:
\begin{align}
H_{\rm stat}(k)=\varepsilon(k)\tau_1-\Delta(k)\tau_2\equiv\begin{pmatrix}0&h(k)\\h^\dag(k)&0
\end{pmatrix},
\end{align}

\noi with the help of a unitary transformation generated by $R=\big(\tau_3+\tau_1\big)/\sqrt{2}$. We find that $h(k)=\varepsilon(k)+i\Delta(k)$.

\begin{figure}[t!]
\begin{centering}
\includegraphics[width=0.494\columnwidth]{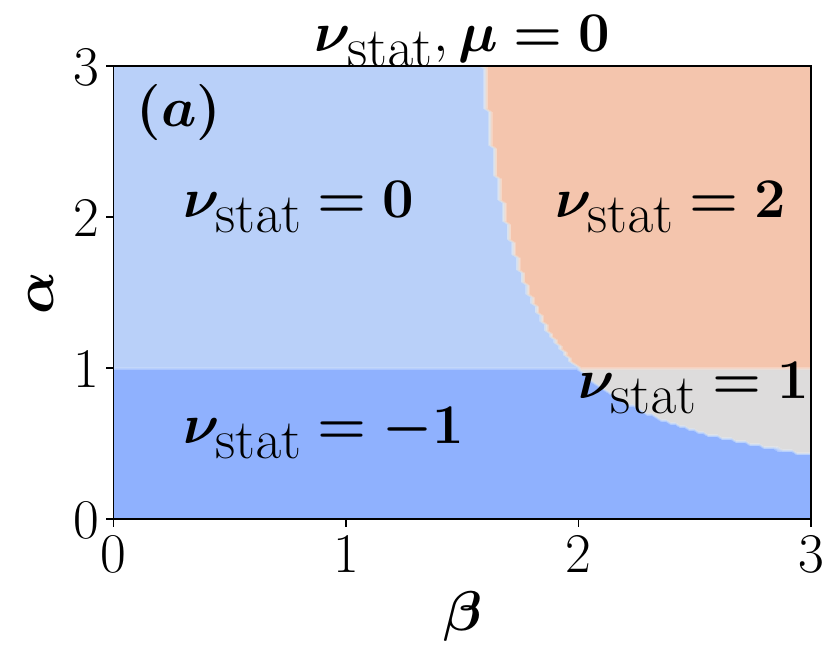}
\includegraphics[width=0.494\columnwidth]{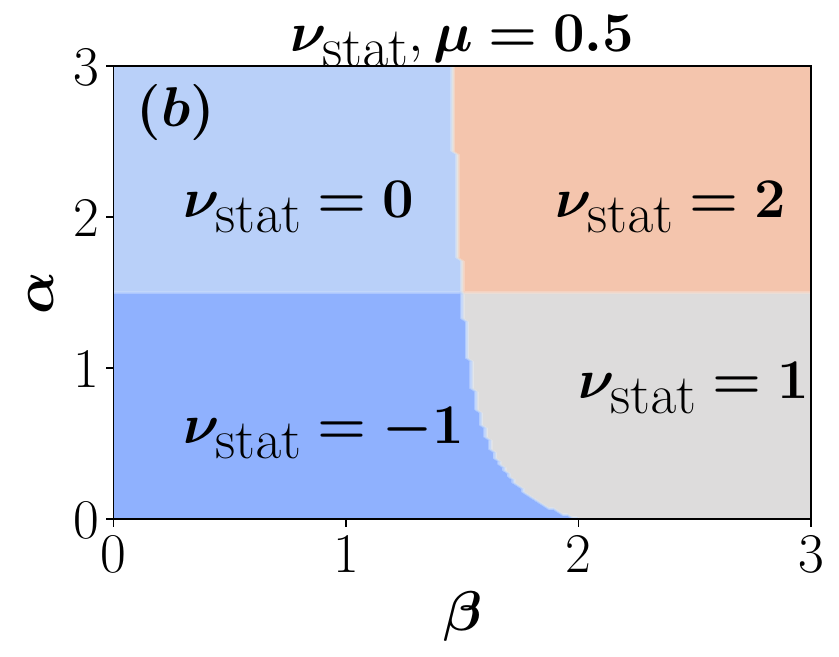}
\end{centering}
\caption{Topological phase diagrams of the static extended Kitaev chain model in Eq.~\eqref{eq:BDIHam} as a function of $\alpha$ and $\beta$, which correspond to the next neighbor hopping and pairing terms. (a) and (b) are obtained for $\mu=0$ and $\mu=0.5$, with $\mu$ being expressed in units of $\Delta$ which is set to unity. These diagrams reveal a variety of topological phases, in which the respective winding number $\nu_{\rm stat}$ takes the values $\{0,\pm1,2\}$.} \label{fig:Figure1}
\end{figure}

The Hamiltonian $H_{\rm stat}(k)$ is of the BDI type since besides the chiral symmetry with $\Gamma=\tau_3$, it additio\-nal\-ly possesses the generalized time-reversal and charge-conjugation symmetries effected by the operators $\Theta=K$ and $\Xi=\tau_3K$, respectively. Here, $K$ denotes the complex conjugation operator. The topological pro\-per\-ties of $H_{\rm stat}(k)$ are described by the winding number~\cite{Ryu2010} $\nu_{\rm stat}=\int_0^{2\pi}d\varphi_{\rm stat}/2\pi$, where $\tan[\varphi_{\rm stat}(k)]=\Delta(k)/\varepsilon(k)$. By numerically evaluating $\nu_{\rm stat}$, we obtain a rich topological phase landscape which is depicted in Fig.~\ref{fig:Figure1}.

We now consider the topological properties of the extended Kitaev model for the driven Hamiltonian:
\begin{align}
H(t,k)=H_{\rm stat}(k)+\Delta_1(t,k)\tau_2-\Delta_2(t,k)\tau_3\,.
\label{eq:DrivenBDI}
\end{align}

\noi In the above, we introduced two different types of driving terms which are defined as:
\bea
\Delta_1(t,k)&=&\Delta\beta_1\sin(2k)\cos(\omega t)/2\,,\label{eq:Delta1}\\
\Delta_2(t,k)&=&\Delta\beta_2\sin(2k)\sin(\omega t)/2\label{eq:Delta2}\,.
\eea

\noi The first (second) term corresponds to an even-(odd)-under-FCC driving perturbation. The Hamiltonian in Eq.~\eqref{eq:DrivenBDI} preserves the symmetries generated by the action of the operators $\Gamma=\tau_3$, $\Theta=K$, and $\Xi=\tau_3K$. Hence, it is also of the BDI type. In fact, since the charge-conjugation symmetry is built-in in the BdG formalism~\cite{AZ}, additional requirement of chiral symmetry imposed throughout this work implies that no symmetry class conversion can occur in BDI systems.

\subsection{Topological phases for even-under-FCC drivings}\label{sec:BDIeven}

We now investigate the topological properties of the extended Kitaev model when $\beta_1\neq0$ while at the same time $\beta_2=0$. Hence, the driving term here reads as $V_1(t,k)={\cal B}_1(k)\omega\cos(\omega t)\tau_2/2$, which is expressed in terms of the function ${\cal B}_1(k)=\big(\Delta\beta_1/\omega\big)\sin(2k)$. Along the spirit of Ref.~\onlinecite{Benito2014}, it is preferrable to employ a time-dependent unitary transformation ${\cal S}_1(t,k;t_0=0)$ which eliminates $V_1(t,k)$ from the Floquet operator\footnote{We transfer to a new frame by re-wri\-ting the Floquet operator according to the form ${\cal H}'(t,k)={\cal S}^\dag(t,k;t_0){\cal H}(t,k){\cal S}(t,k;t_0)={\cal S}^\dag(t,k;t_0)H(t,k){\cal S}(t,k;t_0)-{\cal S}^\dag(t,k;t_0)i\partial_t{\cal S}(t,k;t_0)-i\partial_t\mathds{1}_H$. The latter is re-expressed as ${\cal H}'(t,k)=H'(t,k)-i\partial_t\mathds{1}_H$, with $H'(t,k)={\cal S}^\dag(t,k;t_0)H(t,k){\cal S}(t,k;t_0)-{\cal S}^\dag(t,k;t_0)i\partial_t{\cal S}(t,k;t_0)$. This is achieved by choosing ${\cal S}(t,k;t_0)=
{\rm Exp}\big[-i\int_{t_0}^td\tau\,V(\tau,k)\big]$.}. For the given driving potential, we choose $t_0=0$ and find:
\begin{align}
{\cal S}_1(t,k;t_0=0)=e^{-i\theta_1(t,k)\tau_2/2},
\label{eq:S1}
\end{align}

\noi with the phase $\theta_1(t,k)={\cal B}_1(k)\sin(\omega t)$. In this new frame, the time-dependent Hamiltonian becomes:
\begin{align}
H_1'(t,k)=\varepsilon(k)\left[e^{i\theta_1(t,k)}\frac{\tau_1-i\tau_3}{2}+{\rm h.c.}\right]-\Delta(k)\tau_2,
\label{eq:HamPrime}
\end{align}

\noi The advantage of transferring to a new frame is that the Fourier transform of the Hamiltonian $H'_1(t,k)$ is easier to obtain than for $H_1(t,k)$. Here, Fourier transforming $H'_1(t,k)$ yields the following QEO matrix elements:
\bea
&&{\cal H}_{1;s;n,m}'(k)=\Omega_{s;n,m}+J^{+}_{n-m}\big[{\cal B}_1(k)\big]\varepsilon(k)\tau_1\no\\
&&\quad\,\qquad\qquad-\Delta(k)\tau_2\delta_{n,m}-iJ^{-}_{n-m}\big[{\cal B}_1(k)\big]\varepsilon(k)\tau_3
\qquad\,\label{eq:H1prime}
\eea

\noi where we introduced the functions $
J^{\pm}_n=\big(J_n\pm J_{-n}\big)/2$, with $J_n$ denoting the Bessel functions of the first kind with an order $n\in\mathbb{Z}$. By virtue of the relation $J_{-n}=(-1)^nJ_n$, we conclude that $J^+_n(k)$ ($J^-_n(k)$) are nonzero for an even (odd) $n$. Finally, this property, in conjunction with the fact that $J_n(-k)=\pm J_n(k)$ for an even/odd $n$, respectively, further implies that $J^\pm_n(k)=\pm J^\pm_n(-k)$.

\subsubsection{High-frequency driving regime}\label{sec:BDIHighFreq}

With the QEOs at hand, we proceed with inferring the accessible topological phases in the presence of the dri\-ving. We first discuss the high-frequency limit, in which case, the frequency is much larger than the bandwidth defined according to $W={\rm max}\big[2|E(k)|\big]$. In this high-frequency regime, the zeroth order of the Floquet-Magnus (FM) expansion which is obtained from Eq.~\eqref{eq:H1prime} by taking the time-average over a single period, already constitutes a good appro\-xi\-mation~\cite{Benito2014,li2017,li2018,blanes2009}. Therefore, in this frequency window the driven system can be described by the static FM Hamiltonian $
H_{1;{\rm FM}}'(k)=\tilde{\varepsilon}(k)\tau_1-\Delta(k)\tau_2$. Quite remar\-ka\-bly, this Hamiltonian shares the same structure with $H_{\rm stat}(k)$, albeit with a rescaled normal-phase ener\-gy dispersion given as $\tilde{\varepsilon}(k)=J_0\big[{\cal B}_1(k)\big]\varepsilon(k)$ which tends to $\varepsilon(k)$ in the limit $\omega\rightarrow\infty$.

\begin{figure}[t!]
\begin{center}
\includegraphics[width=0.494\columnwidth]{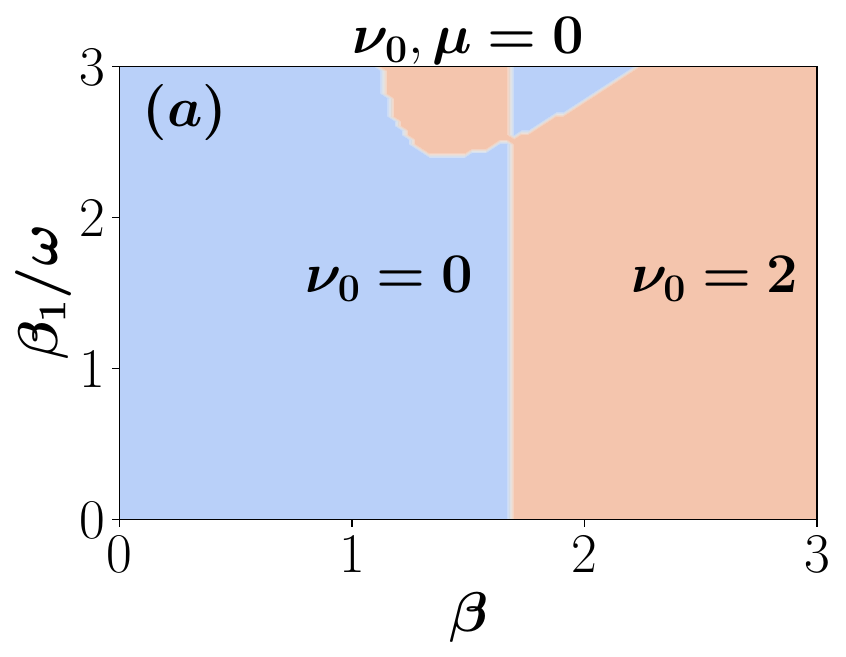}
\includegraphics[width=0.494\columnwidth]{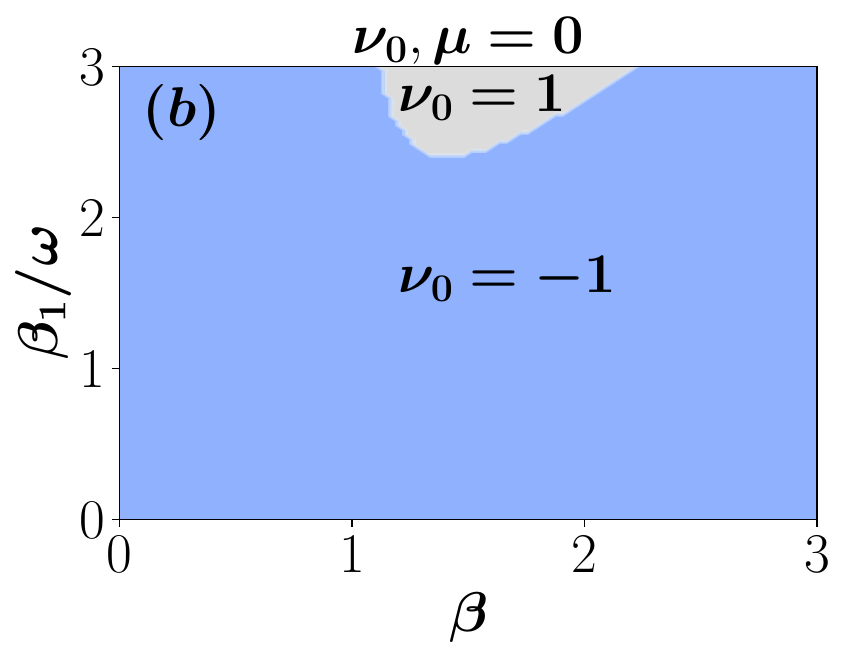}
\end{center}
\caption{Topological phase diagrams of the driven extended Kitaev chain model for a driving perturbation which is even under the action of FCC. The diagrams are obtained in the high frequency regime, and are depicted as functions of the parameter $\beta$ and the reduced drive amplitude $\beta_1/\omega$. Panels (a) and (b) illustrate the topological invariants $\nu_0$ using the zeroth order Floquet-Magnus approach for $\alpha=2$ and $\alpha=0$, respectively. We used $\mu=0$ and $\Delta=1$ in both panels. The bandwidth for $\alpha=2$ is $W=6$. The results of panel (a) are also confirmed using the QEO approach for $\omega=10$. Moreover, using the QEO approach, we find that $\nu_\pi=0$ in the same parameter window. This is consistent with the expected absence of Majorana $\pi$ modes for frequencies which are larger than the energy bandwidth of the static system.}
\label{fig:Figure2}
\end{figure}

The FM Hamiltonian describes the topological pro\-per\-ties for the $\varepsilon=0$ quasi-energy, since MPMs are not accessible in this limit. Fi\-gu\-res~\ref{fig:Figure2}(a) and~\ref{fig:Figure2}(b) depict the topological phase diagrams obtained using the zeroth order FM method for two different values of the parameter $\alpha$. Our FM results are further confirmed by numerically evaluating the Floquet invariants for a finite high frequency value. Indeed, by setting $\omega=10>W$, we verify the good agreement of the two methods.

\subsubsection{Low-frequency driving regime}\label{sec:BDILowFreq}

In the lower frequency regime, where the bandwidth $W$ is larger than the driving frequency $\omega$, the driven system deviates from the static-like behavior observed above. That is, now both MZMs and MPMs are simultaneously accessible. In contrast to the FM method, these effects are fully captured by the QEO approach. We remark that, throughout this work, in order to carry out all the numerical calculations using the QEO approach, we truncated the space spanned by the operator $\check{N}$ down to the subspace $\check{N}=\{1,2\}$.

\begin{figure}[t!]
\begin{center}
\includegraphics[width=0.494\columnwidth]{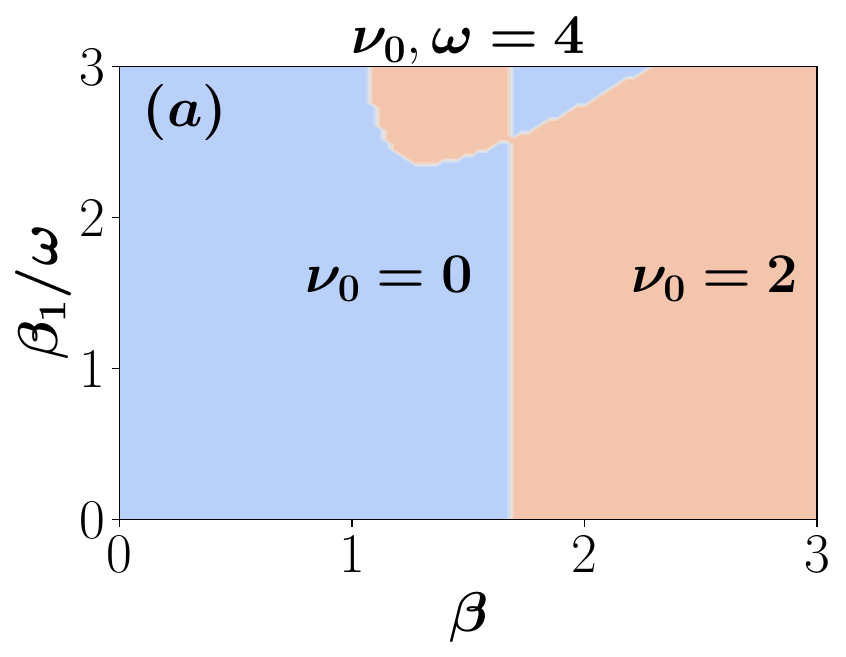}
\includegraphics[width=0.494\columnwidth]{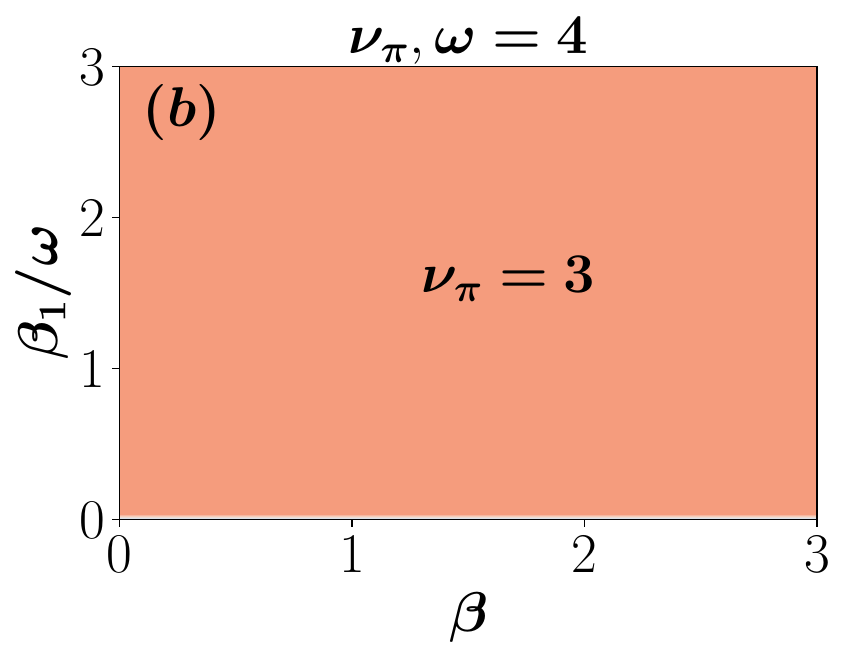}\\
\includegraphics[width=0.494\columnwidth]{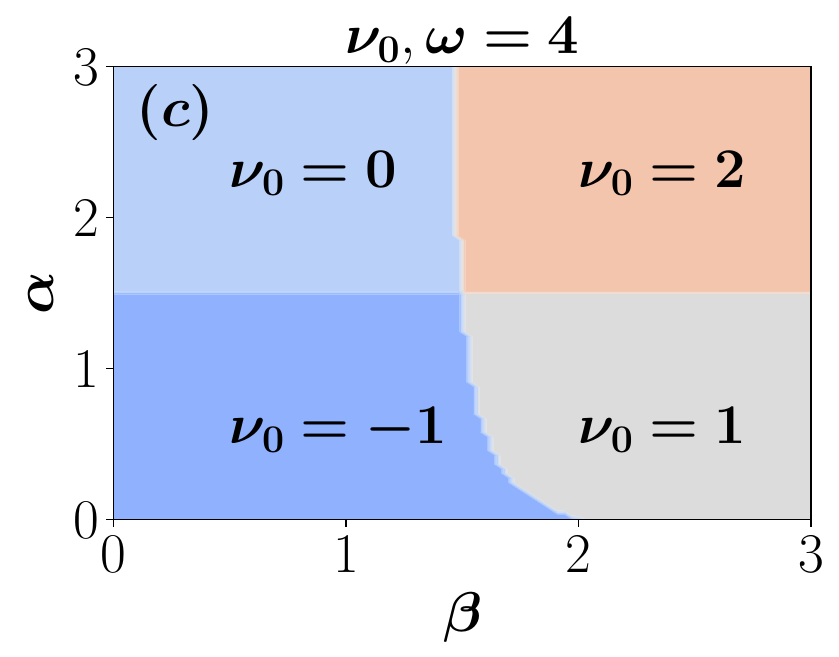}
\includegraphics[width=0.494\columnwidth]{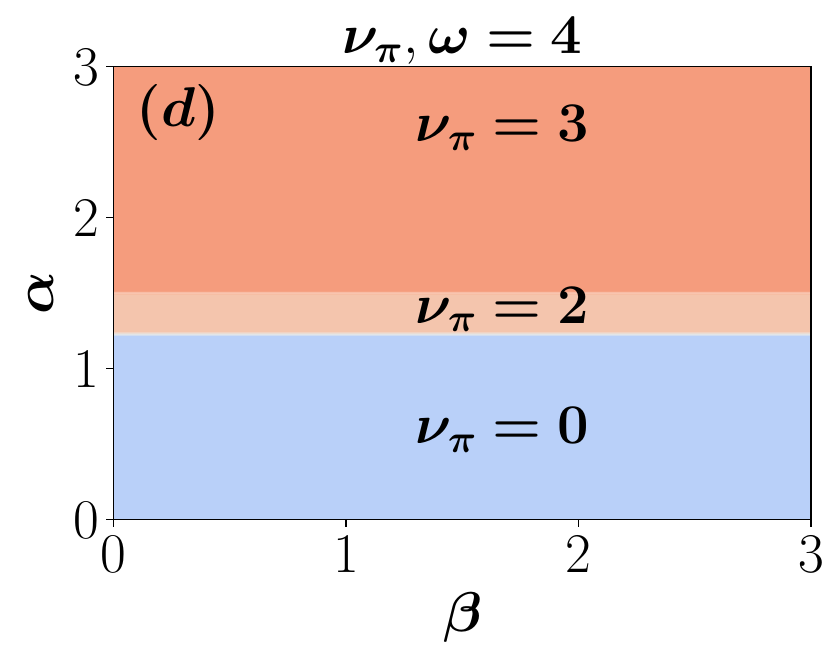}
\end{center}
\caption{Numerical evaluation of the topological phase diagrams associated with the winding numbers $\nu_{0,\pi}$ as a function of $\alpha$, $\beta$, and the reduced drive amplitude $\beta_1/\omega$. The results of panels (a)-(b) were obtained using the QEO method for $\mu=0$ and $\alpha=2$. Here, we consider the low-frequency regime with $\omega=4$ and a bandwidth $W=6$. In comparison to the results in Fig.~\ref{fig:Figure2}(a), which were obtained using the zeroth order Floquet-Magnus approach, we find that reducing the driving frequency practically leaves the emergence of MZMs unaffected. More importantly, we find that switching on the driving immediately induces three MPMs per edge. In panels (c) and (d) we fix the reduced drive amplitude to $\beta_1/\omega=0.5$ and the chemical potential to $\mu=0.5$, and vary $\alpha$ and $\beta$. The results in (c) are very similar to the ones of Fig.~\ref{fig:Figure1}(a) in this regime. We finally note that $\nu_\pi$ demonstrates substantial variations upon increasing $\alpha$. This is because this variable effectively sets the value of the bandwidth and thus controls whether the static energy bands cross the boundaries of the first Floquet zone and, in turn, undergo gap openings.}
\label{fig:Figure3}
\end{figure}

Our numerical results are shown in Figs.~\ref{fig:Figure3}(a)-(b). Notably, we find that the system also harbors multiple MPMs per edge. In more detail, by setting $\omega=4$, we find that $\nu_\pi=3$ in the entire $\big(\beta,\beta_1/\omega\big)$ pa\-ra\-me\-ter plane. This is in accordance with the general trend found in the analysis of Sec.~\ref{sec:piEvenFCCanalytics}, i.e., that MPMs become accessible even at infinitesimally small drive amplitudes. Moreover, in Figs.~\ref{fig:Figure3}(c)-(d) we also show the resulting topological phase diagrams in the $(\alpha,\beta)$ plane for $\mu=0.5$ and a reduced drive amplitude $\beta_1/\omega=0.5$. Figure~\ref{fig:Figure3}(c) appears very similar to Figure~\ref{fig:Figure1}(a), therefore implying that for weak drive amplitudes the $\nu_0$ sector is practically left unmodified. In contrast, the $\nu_\pi$ undergoes changes which correspond to topological phase transitions leading to multiple MPMs per edge.

\subsubsection{Comparing QEO and TEO methods}\label{sec:BDIComparison}

As we have already pointed out, the QEO method is expected to be more suitable for tackling the problems discussed here compared to previously employed TEO approaches~\cite{kitagawa2010b,Asboth2014}, because of the harmonic nature of the driving. To provide support for this claim, we have also attempted to reproduce the results of Fig.~\ref{fig:Figure3}(a)-(b) by employing the half-period TEO approach~\cite{Asboth2014}. Details of this exploration are presented in Appendix~\ref{sec:TEOvsQEO}. Indeed, as we discuss there, the TEO approach requires at least an order of magnitude more computational runtime, in order to arrive at the same numerical accuracy obtained for the topological inva\-riants using the QEO method.

By means of the TEO approach, one also is in a position to obtain the so-called effective Hamiltonian~\cite{eckardt2015}. Spe\-ci\-fi\-cal\-ly, this is an effectively static Hamiltonian defined from the Floquet operator, or, the stroboscopic TEO. Hence, to complete the study of the topological properties in terms of the TEO approach, in Appendix~\ref{sec:BandInvariants} we also evaluate the TEO for a full period and, in turn, obtain the effective Hamiltonian. By means of the latter, we evaluate the so-called band inva\-riants and identify their relation with the $\nu_{0,\pi}$ invariants.

We note that we restrict the comparison between the QEO and TEO approaches only to the present case, since similar results are also expected for rest of the scenarios that we examine later on.

\subsection{Topological phases for odd-under-FCC drivings}\label{sec:BDIodd}

We now proceed and examine the case for which $\beta_2\neq 0$ and $\beta_1=0$. Here, the driving term becomes $V_2(t,k)=-{\cal B}_2(k)\omega\sin(\omega t)\tau_3/2$, where we set ${\cal B}_2(k)=\big(\Delta\beta_2/\omega\big)\sin(2k)$. The symmetry class BDI persists also for this driving term. Following the same procedure as in the previous section, one can transfer to a new Hamiltonian frame by means of a unitary transformation effected now by the operator:
\begin{align}
{\cal S}_2(t,k;t_0=\pi/2)=e^{-i\theta_2(t,k)\tau_3/2},
\end{align}

\noi with $\theta_2(t,k)={\cal B}_2(k)\cos(\omega t)$. As a result, the time-dependent Hamiltonian in the new frame takes the form:
\begin{align}
H_2'(t,k)=E(k)\left\{e^{i[\varphi_{\rm stat}(k)+\theta_2(t,k)]}\tau_++{\rm h.c.}\right\},
\label{eq:HamPrime2}
\end{align}

\noi where we introduced the matrices $\tau_\pm=(\tau_1\pm i\tau_2)/2$, while we made use of the variables $E(k)$ and $\varphi_{\rm stat}(k)$ defined in Sec.~\ref{sec:BDI}. Once again, it is straightforward to obtain the Fourier transform for the above Hamiltonian. After carrying this calculation out, the resulting expressions for the QEOs read as follows:
\bea
&&{\cal H}_{2;s;n,m}'(k)=\Omega_{s;n,m}+E(k)\Big\{\no\\
&&\qquad\quad\qquad i^{n-m}J_{n-m}\big[{\cal B}_2(k)\big]e^{i\varphi_{\rm stat}(k)}\tau_++{\rm h.c.}\Big\}.\qquad
\label{eq:H2prime}
\eea

\begin{figure}[t!]
\begin{center}
\includegraphics[width=0.494\columnwidth]{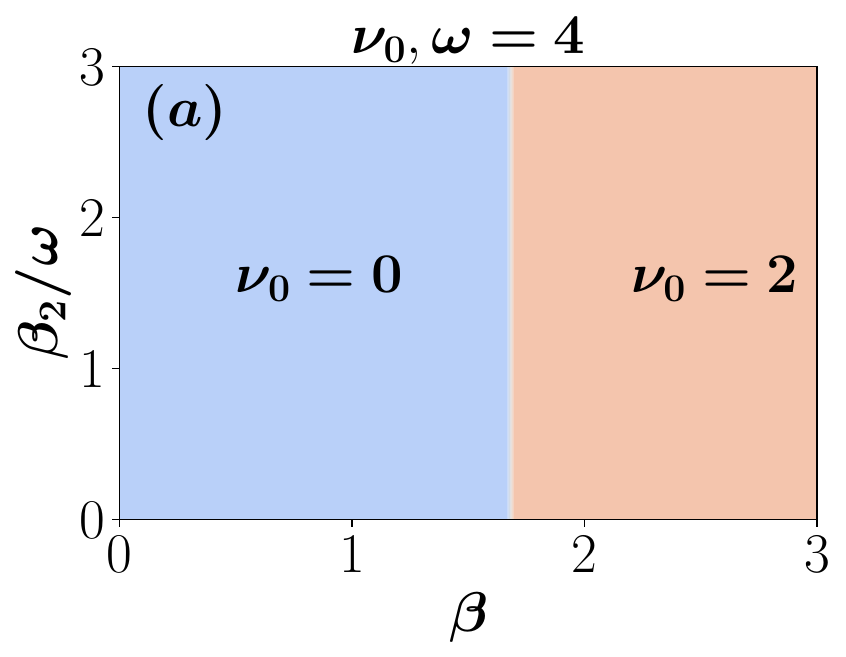}
\includegraphics[width=0.49\columnwidth]{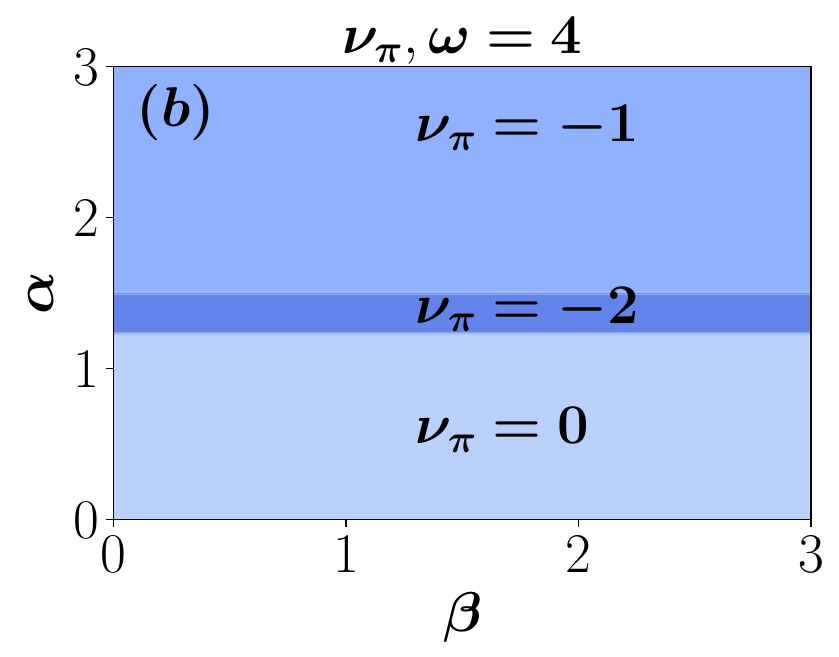}
\end{center}
\caption{Topological phase diagrams of the driven extended Kitaev chain model in Eq.~\eqref{eq:BDIHam} for an odd-under-FCC dri\-ving perturbation. In (a) the topological phase diagram is depicted as a function of $\beta$ and the reduced drive amplitude $\beta_2/\omega$, when $\mu=0$, $\Delta=1$, $\alpha=2$, and $W=6$. In contrast, in (b) we explore the dependence of the winding number $\nu_\pi$ on $\alpha$ and $\beta$. Here, $\mu=0.5$, $\beta_2/\omega=0.5$ and all other parameter values remain unchanged. In (a) we observe that the phase diagram is not influenced by the drive. Note that the same diagram is obtained for two different frequency values $\omega=\{4,10\}$. Most importantly, (a) does not differ from the results of the static case. This is consistent with the discussion in Sec.~\ref{sec:Heuristic} and Eq.~\eqref{eq:TrendsZeroOdd}. Even more, the corresponding topological phase diagram for $\omega=4$ in the plane $(\alpha,\beta)$ is identical to Fig.~\ref{fig:Figure3}(c). Although the topological properties at zero quasi-energy are not affected by the drive,  MPMs are generally accessible. Specifically, for the case in panel (a) we correspondingly find $\nu_{\pi}=0$ and $\nu_{\pi}=-1$ for $\omega=10$ and $\omega=4$. In fact, for the latter case, MPMs immediately emerge for as soon as the drive amplitude becomes nonzero, since the frequency is smaller than the bandwidth. This property is consistent with the expectations from Eq.~\eqref{eq:TrendsPiOdd}. Panel (b) shows that the $\nu_\pi$ inva\-riant is affected in a radically different manner than the one in Fig.~\ref{fig:Figure3}(d) obtained for an even-under-FCC drive with amplitude $\beta_1\neq0$.}
\label{fig:Figure4}
\end{figure}

From the above expression, we find that in the high-frequency regime, the zeroth-order FM Hamiltonian is given as $H_{2;{\rm FM}}'(k)=J_{0}\big[{\cal B}_2(k)\big]H_{\rm stat}(k)$. This analytical result aligns well with corresponding numerical results obtained using the QEO approach which are discussed in Fig.~\ref{fig:Figure4}. Furthermore, our numerical results obtained through the QEO approach, illustrate the emergence of MZMs and MPMs in the lower frequency regime. Notably, in both regimes, the MZMs exhibit negligible deviations from their counterparts in the even-under-FCC driving scenario, cf Fig.~\ref{fig:Figure3}. Indeed, the phase diagrams  of the static and driven systems at zero quasi-energy are essentially identical. For MPMs, the to\-po\-lo\-gical phases diagrams manifest sensitivity to the driving mechanism and the number of edge states are different from those inferred for an even-under-FCC drive. See also Fig.~\ref{fig:Figure3}(d).

\section{Analysis of a representative CI model in 1D}\label{sec:CI}

We now move on with considering that our static Hamiltonian belongs to class CI instead of BDI. Such a symmetry class change can be obtained by assuming the same structure for the static Hamiltonian, that is:
\begin{align}
H_{\rm CI}(k)=\varepsilon(k)\tau_3+\bar{\Delta}(k)\tau_2\,,
\label{eq:CI}
\end{align}

\noi where the energy dispersion $\varepsilon(k)$ remains unchanged, but with the difference that now the static pairing gap is given by $\bar{\Delta}(k)=\Delta\cos k$. This modification is sufficient to enlist the present Hamiltonian in the CI symmetry class. Notably, such a Hamiltonian can describe a 1D spin-singlet superconductor with a so-called extended s-wave pairing gap~\cite{Mazin}. In this case, the $\tau_{1,2,3}$ Pauli matrices are defined in a Nambu space spanned by the following electron $\big|e;k,\uparrow\big>$ and hole states $\big|h;-k,\downarrow\big>$, which correspond to the eigenstates $\tau_3=\pm1$, respectively. This two-component Nambu formalism is capable of descri\-bing superconductors which preserve rotations about the spin quantization axis and support Cooper pairs with zero net spin angular momentum~\cite{SigristUeda}. Hence, besides even-parity spin-singlet superconductivity, also the odd-parity spin-triplet zero-spin pairing  state can be simultaneously described using the above spinor.

To proceed, we bring the above Hamiltonian in a block off-diagonal form by means of the unitary transformation $\bar{H}_{\rm stat}(k)=R^\dag H_{\rm CI}(k)R=\varepsilon(k)\tau_1-\bar{\Delta}(k)\tau_2$. This Hamiltonian possesses a chiral symmetry with $\Gamma=\tau_3$, a gene\-ra\-li\-zed time-reversal symmetry with $\Theta=\tau_1K$, and lastly, a charge-conjugation symmetry with $\Xi=i\tau_2K$. The topological properties of this Hamiltonian are obtained from the winding number of the phase $\bar{\varphi}_{\rm stat}(k)$ which is defined through the relation $\tan\big[\bar{\varphi}_{\rm stat}(k)\big]=\bar{\Delta}(k)/\varepsilon(k)$. In 1D, Hamiltonians in the CI symmetry class do not harbor topologically nontrivial phases~\cite{Ryu2010}. For the Hamiltonian in Eq.~\eqref{eq:CI}, this is because the Berry singularities appea\-ring at $k=\pm\pi/2$ carry opposite charges. As a result, they cancel each others contribution and lead to a zero winding number for $\bar{\varphi}_{\rm stat}(k)$.

As we discussed in Sec.~\ref{sec:NewScenarios}, however, the addition of the drive can allow for the symmetry class conversion CI$\rightarrow$AIII and, in turn, enable the system to support topo\-lo\-gi\-cal\-ly nontrivial phases. Such a symmetry class conversion scenario becomes possible, for instance, by considering the following driven Hamiltonian:
\begin{align}
\bar{H}(t,k)=\bar{H}_{\rm stat}(k)+\Delta_1(t,k)\tau_2-\Delta_2(t,k)\tau_3\,,
\end{align}

\noi where the time-dependent pairing gaps are given once again by the formulas in Eqs.~\eqref{eq:Delta1} and~\eqref{eq:Delta2}. Note that, in the Nambu basis of the Hamiltonian in Eq.~\eqref{eq:CI}, either driving terms describes a time-dependent odd-parity spin-triplet gap where each Cooper pair lies in the spin-symmetric spin-zero state $\propto\big|k,\uparrow;-k\downarrow\big>+\big|k,\downarrow;-k\uparrow\big>$.

{\color{black} It is important to note that the symmetry class conversion CI$\rightarrow$AIII can also take place under fully-static conditions. In fact, by modifying $\bar{\Delta}(k)$ so that it also includes pairing terms which are odd in $k$, effects the above transition and opens the door to nontrivial topological edge modes at zero energy. However, the dy\-na\-mi\-cal control of such a type of conversion that we inve\-sti\-gate and propose here has advantages when it comes to applications, since it can unlock the on demand ge\-ne\-ra\-tion of boundary modes and other features unique to driven systems. Evenmore, by employing driving to mediate this transition renders us in a position to choose the characteristic energy at which the topological edge modes appear, hence enhancing the control over the topological properties of the system.}

To this end, we remark that considering the same type of drives as for the BDI Hamiltonian studied earlier, allows us to take advantage of the various analytical formulas obtained in Sec.~\ref{sec:BDI}. In this manner, the exploration of the topological properties of the driven CI Hamiltonian becomes significantly facilitated, while at the same time, this allows for a direct comparison between the effects of dri\-ving on Hamiltonians belonging to the two different symmetry classes.

\subsection{Topological phases for even-under-FCC drivings}\label{sec:CIeven}

We first discuss the case $\beta_1\neq0$ and $\beta_2=0$. In the same spirit of Sec.~\ref{sec:BDIeven}, we perform a unitary rotation and obtain the corresponding QEO, which has an identical form to the QEO in Eq.~\eqref{eq:H1prime}, albeit with the static pairing gap $\Delta(k)$ replaced by $\bar{\Delta}(k)$.

\begin{figure}[t!]
\begin{center}
\includegraphics[width=0.494\columnwidth]{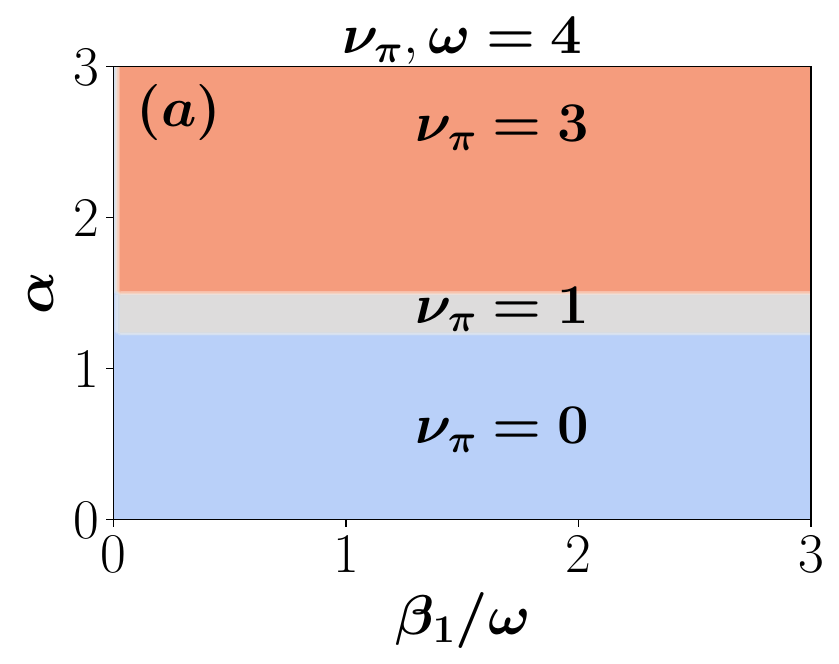}
\includegraphics[width=0.494\columnwidth]{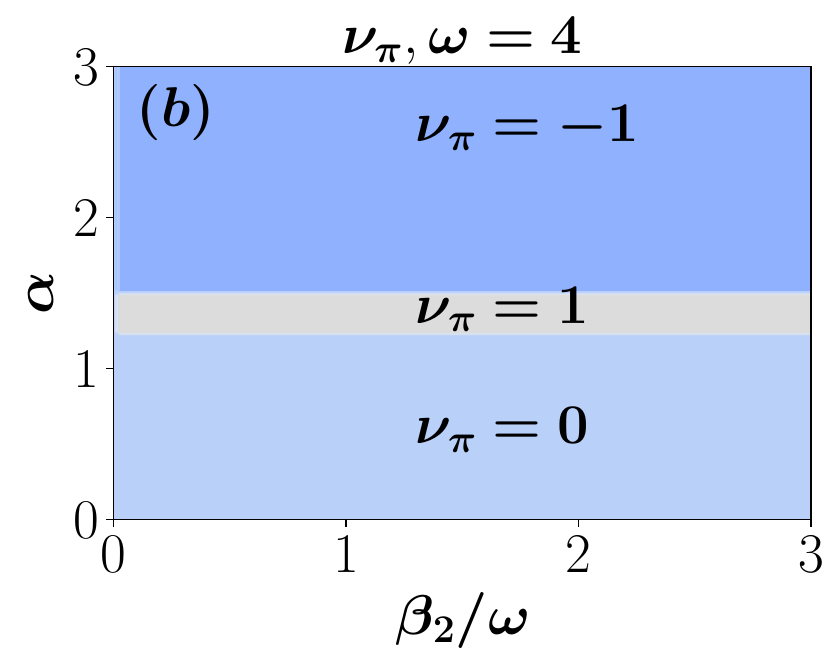}
\end{center}
\caption{Panels (a) and (b) present the numerical evaluation of the topological phase diagrams associated with the winding number $\nu_\pi$ as a function of $\alpha$ and the reduced drive amplitude $\beta_1/\omega$ and $\beta_2/\omega$, respectively. These are both obtained in the low frequency regime with $\omega=4$ and a chemical potential value $\mu=0.5$. The above results describe the driving-induced symmetry class conversion CI$\rightarrow$AIII. We observe that in contrast to the topologically-trivial static CI Hamiltonian, the driven system shows various phases harboring $\pi$ modes. Interestingly, the one can obtain a different topological landscape depending on the time of driving employed. Finally, we have checked that in the high drive frequency regime ($\omega=10$) $\nu_{0,\pi}=0$, while $\nu_0$ remains zero also for $\omega=4$.}
\label{fig:Figure5}
\end{figure}

With the help of the QEO approach we obtain the respective topological invariants $\nu_{0,\pi}$ for both high ($\omega=10$) and low ($\omega=4$) frequencies. In the former case, we find $\nu_{0,\pi}=0$ in accordance with the expected topologically trivial behavior of a system in CI symmetry class. By instead swit\-ching to a low-frequency driving protocol, topologically nontrivial phases become now accessible but only for the $\pi$ sector. Indeed, $\nu_0$ is identically zero even for $\omega=4$ for the entire range of parameter values of $\alpha$, $\beta$, and $\beta_1/\omega$ typically examined in this work. Due to the triviality of $\nu_0$, we only show results for $\nu_\pi$. Specifically, in Fig.~\ref{fig:Figure5}(a) we depict the evolution of $\nu_\pi$ as a function of $\alpha$ and $\beta_1/\omega$. We find that $\nu_\pi$ takes a variety of values which strongly depend on the value of $\alpha$, which sets the energy bandwidth $W$ of the static system.

Concluding this section, it is important to emphasize that the topologically protected zero and $\pi$ quasi-energy edge modes which are predicted by $\nu_{0,\pi}$ for a Hamiltonian in the AIII symmetry class do not correspond to Majorana excitations. This becomes transparent from the fact that the Nambu space in the present case consists of electrons and holes of opposite spins, therefore, charge neutral excitations are inaccessible, see also Ref.~\onlinecite{KotetesClassi}. In stark contrast, the ari\-sing topological excitations can be viewed as topo\-lo\-gi\-cal\-ly pro\-tec\-ted Andreev modes. A number of works have discussed the emergence of topological Andreev modes but, up to our knowledge, so far only in static systems~\cite{KlinovajaFF,Rainis,SticletFF,NagaosaFF,MarraFF,Steffensen2022}. Specifically, Refs.~\cite{KlinovajaFF,Rainis,SticletFF,NagaosaFF,MarraFF} have discussed zero-energy topological Andreev modes in a parameter space, while Ref.~\onlinecite{Steffensen2022} has proposed a mecha\-nism to induce such zero modes at the ends of a 1D superconductor. Therefore, our results reveal alternative paths to engineer topological Andreev zero and $\pi$ modes in driven systems.

\subsection{Topological phases for odd-under-FCC drivings}\label{sec:CIodd}

We now proceed with the last topological scenario to be examined in this work, that is $\beta_1=0$ and $\beta_2\neq0$. In analogy to the previous paragraph, we now follow the same steps as in Sec.~\ref{sec:BDIodd} and find the QEO in the present case, which has the same structure with the one in Eq.~\eqref{eq:H2prime} but with the pairing gap properly substituted.

Our numerical calculations reveal that also here, Andreev zero modes are not accessible, at least not for the parameter window typically examined throughout this work. Andreev $\pi$ modes are instead relevant. In Fig.~\ref{fig:Figure5}(b) we show the resulting topological phase diagram in the $\big(\beta_2/\omega,\alpha\big)$ parameter plane. Interestingly, we observe that for high values of $\alpha$, the $\pi$ phases stabilized for even and odd drivings under FCC are substantially different. Hence, the phase offset of the harmonic driving matters, and can constitute a topological control knob. The resulting difference can be corroborated after the discussion of Sec.~\ref{sec:Heuristic}. In the case of the odd-under-FCC driving, the topological phase transition $\nu_\pi=-1\mapsto\nu_\pi=+1$ stems from two simultaneous gap closings of the driving potential. In contrast, for the even-under-FCC case, the driving leads to a more complex interplay of the momentum space structure of the {\color{black}driving} perturbation and the static Hamiltonian, thus, leading to a different result.

\section{Conclusions}\label{sec:Conclusions}

We explore the topological properties of FTIs posses\-sing a DCS. Since our focus is on harmonic drivings, we employ the QEO technique.  We identify the chiral symmetry operator in frequency space and discuss the construction of the topological invariants. Specifically, we determine a closed analytical expression for the operator that yields the bulk invariants $\nu_{0,\pi}$. These predict the emergence of topological boundary modes at zero and $\pi$ quasi-energy, respectively. With these general framework, we analyze the topological properties of FTIs with DCS, and distinguish two distinct scenarios depending on whether the DCS has a static analog or not. We find that in the latter case, the driving tends to have a negligible effect on the topological properties at zero quasi-energy. In contrast, it has a strong impact on the $\pi$-mode to\-po\-lo\-gy, thus opening perspectives for MPM engineering.

We confirm these general trends by applying the QEO method on a concrete extended Kitaev chain model subject to a periodic drive. Our work also verifies that the QEO approach is indeed more efficient for periodic drives than the TEO method for half a period. In addition, the QEO method allows for analytical conclusions and a more transparent discussion of the underlying mechanism for nontrivial topology. Apart from the above mentioned BDI class Kitaev model, we further explore the effects of driving which preserve a DCS while allowing for a symmetry class conversion from a CI to an AIII class. For this demonstration we employed a CI static Hamiltonian which describes an extended s-wave 1D superconductor. This Hamiltonian is topologically trivial in this dimensionality. However, upon switching on a suitable driving perturbation, one can leave only the DCS intact so for the system to transit to the AIII class, which harbors topological modes in 1D. To achieve this symmetry class conversion, we here consider the additional presence of a time-dependent zero-spin spin-triplet pairing.

All in all, our work aims at further expanding on the systematic analysis of driven systems with chiral symmetry using the QEO approach. In addition, it discusses the dichotomy between the two possibilities for the DCS ope\-ra\-tor and illustrates the advantages of employing DCS-preserving drivings without static analog for engi\-nee\-ring topological $\pi$-modes while leaving the zero quasi-energy sector unaffected. Finally, the concrete models that we discussed could in principle be engineered in hybrids of superconductor and semiconducting systems which experience longer-ranged couplings~\cite{kotetes2022}, or, in topological Yu-Shiba-Rusinov systems~\cite{Stevan,Nakosai,Pientka}.

\section*{Acknowledgements}

We are thankful to Frederik Nathan for help\-ful discussions. In addition, we acknowledge funding from the National Na\-tu\-ral Science Foundation of China (Grant No.~12050410262).

\appendix

\section{Comparing QEO and TEO methods}\label{sec:TEOvsQEO}

In this section, we reinfer the low-frequency to\-po\-lo\-gi\-cal phase dia\-grams of Fig.~\ref{fig:Figure3}(a)-(b) using the TEO approach discussed previously in Ref.~\onlinecite{Asboth2014}. Our aim is to compare its computational efficiency to the one of the QEO method. The topological invariants in the TEO method are eva\-lua\-ted by taking advantage of a special structure that the TEO ${\cal U}(t,k)$ obtains at $t=T/2$ for systems with chiral symmetry. Specifically, after expressing the ope\-ra\-tor ${\cal U}(t,k)$ in the basis in which the DCS operator $\Gamma$ becomes block diagonal, we find that:
\begin{align}
{\cal U}(T/2,k)={\cal T}_t\,e^{-i\int_0^{T/2}dt\,H(t,k)}=\begin{pmatrix}
a_\pi(k)&a_0(k)\\b_0(k)&b_\pi(k)
\end{pmatrix}
\end{align}

\begin{figure}[t!]
\begin{center}
\includegraphics[width=0.494\columnwidth]{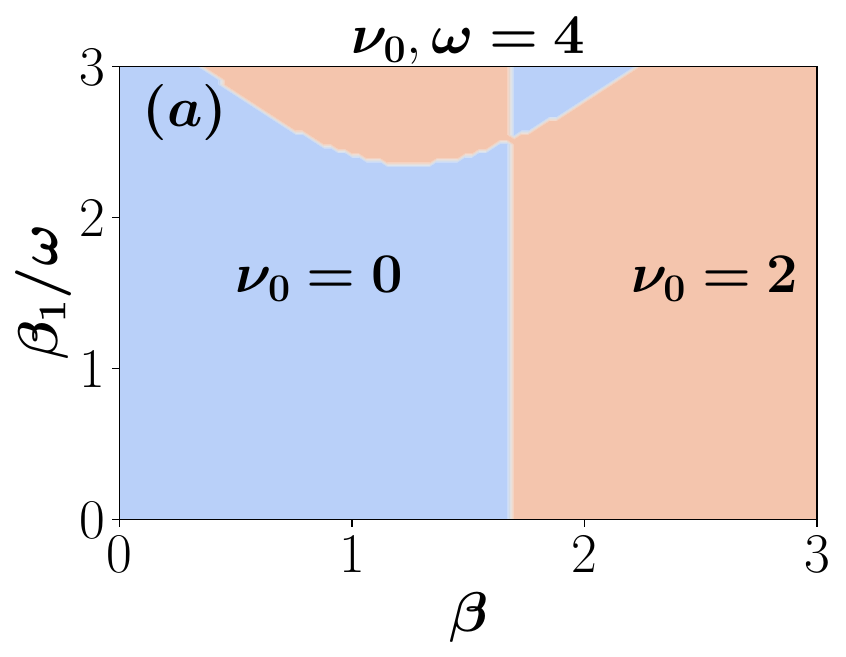}
\includegraphics[width=0.494\columnwidth]{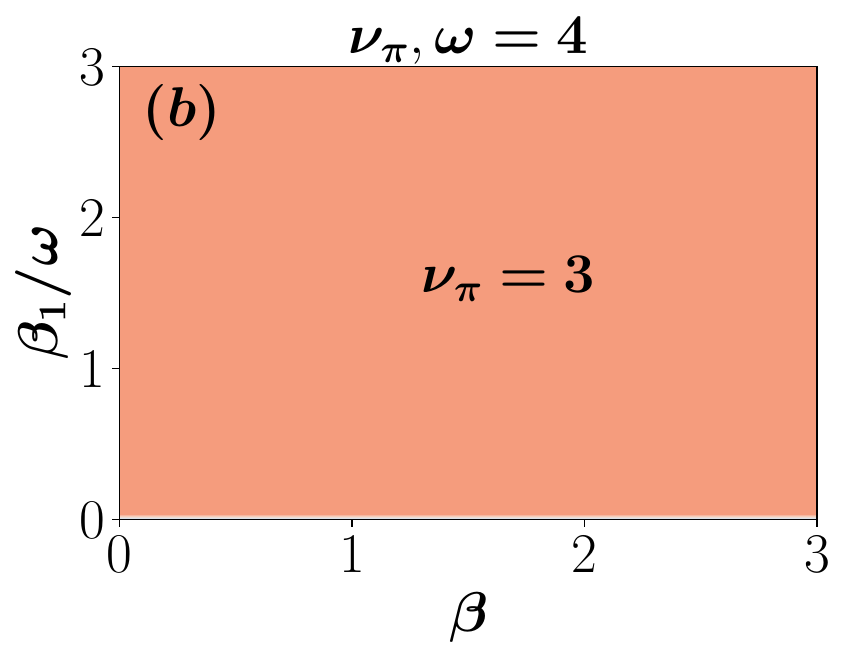}
\end{center}
\caption{Numerical calculation of $\nu_{0,\pi}$ in the parameter plane $(\beta,\beta_1/\omega)$ using the TEO approach. We use $\Delta=1$, $\mu=0$, $\alpha=2$, $\omega=4$, and a number of $M=20$ di\-scretization steps. While the two approaches tend to converge, the QEO is computationally more efficient than its TEO counterpart due to the linear form of the equations in the frequency domain. We have checked that Fig.~\ref{fig:Figure3}(a) is more accurate than Fig.~\ref{fig:Figure1app}(a) by verifying that the former remains unaltered when enlarging the $\check{N}$-space which sets the order of truncation for the QEO.}
\label{fig:Figure1app}
\end{figure}

\noi where ${\cal T}_t$ denotes the time-ordering operator. Within this framework, the invariants $\nu_{0,\pi}$ are identified with the winding numbers of the functions $a_{0,\pi}(k)$, respectively.

In the case of harmonic drivings ${\cal U}(T/2,k)$ is more conveniently evaluated numerically by means of the Suzuki-Trotter decomposition~\cite{trotter1959,Suzuki}. Hence, ${\cal U}(t,k)$ can be approximated as ${\cal U}(t,k)=\prod_{j=0}^{M-1}{\cal U}(t_j+\delta t,t_j;k)$, with ${\cal U}(t_j+\delta t,t_j;k)={\rm exp}\big[-iH(t_j,k)\delta t\big]$, where we set $\delta t=T/M$ and $t_j=j\delta t$. The variable $M$ corresponds to the number of discretization steps~\cite{trotter1959,Suzuki}. The results obtained from applying the above this method to Eq.~\eqref{eq:HamPrime} for the same pa\-ra\-meters as in Fig.~\ref{fig:Figure3}(a)-(b), are depicted in Fig.~\ref{fig:Figure1app}(a)-(b) and are carried out for $M=20$. Notably, we observe that for the latter choice of discretization steps, the results of the TEO method do not yet manage to capture the numerical accuracy that we ended up with by employing the QEO approach.

Aside from the reduced accuracy found when emplo\-ying the TEO method for $M=20$, it is equally important to here comment on the computational efficiency of the two techniques. {\color{black}We indeed find that, for the cases exa\-mi\-ned in this work, the numerical calculation for the QEO is much faster than its TEO counterpart. This should come as no surprise, since the systems of focus belong to the category of harmonically driven systems. Indeed, as it has already been previously discussed~\cite{lindner2013}, for drivings with a frequency decomposition that consists of a small number of frequencies it is preferrable to employ the QEO approach. This is thanks to the localization of the Floquet eigenstates in frequency space, a phenomenon which is analogous to the Wannier-Stark localization taking place in real space. Hence, the harmonic type of driving in conjunction with the legitimate truncation of the QEO operator down to a small number of Floquet zones, are the main culprits for the improved efficiency of the QEO compared to the TEO method observed here. In the present work, our QEO approach has further benefited from an additional factor. This concerns the transformation of the original time-dependent Hamiltonians into a new frame of reference which eli\-mi\-na\-tes the driving term. See for instance the discussion in Sec.~\ref{sec:BDIeven}. This ``trick" allows to re-write the time-dependent Hamiltonians in forms which possess simple Fourier transforms, thus, further contributing to the numerical efficiency of the QEO method.

\begin{figure}[t!]
\begin{center}
\includegraphics[width=0.494\columnwidth]{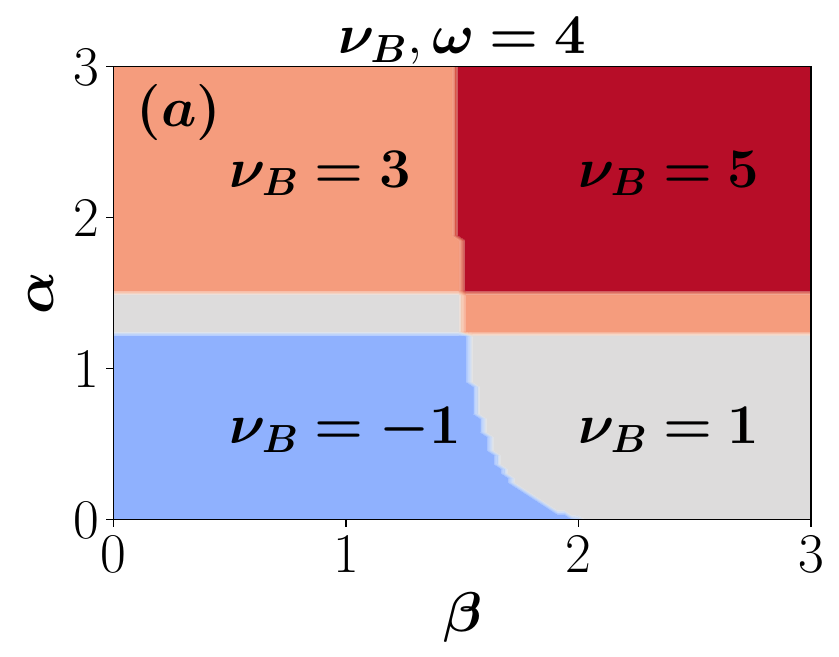}
\includegraphics[width=0.494\columnwidth]{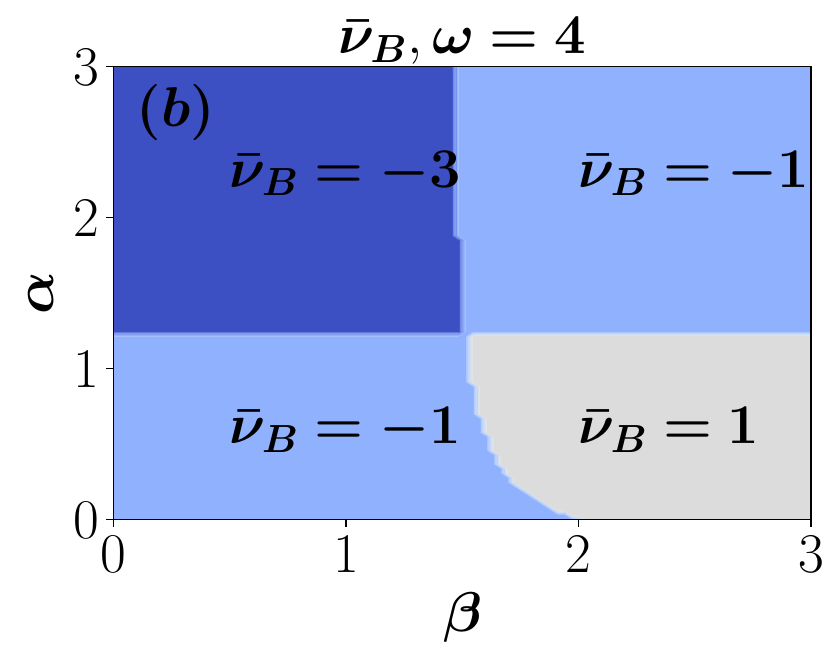}
\end{center}
\caption{Relation between the Floquet invariants $\nu_{0,\pi}$ of Fig.~\ref{fig:Figure3}(c)-(d) and the band invariants $\nu_B$ and $\bar{\nu}_B$ shown in Fig.~\ref{fig:Figure2app}(a)-(b). The band invariants are evaluated using two different time frames separated by half a period. We find that these satisfy the relations $\nu_B=\nu_0+\nu_\pi$ and $\bar{\nu}_B=\nu_0-\nu_\pi$. The calculations were performed for the Hamiltonian in Eq.~\eqref{eq:HamPrime} and the same parameters as in Fig.~\ref{fig:Figure3}(c)-(d).}
\label{fig:Figure2app}
\end{figure}

To quantify the above, we record the computational runtime required to evaluate the phase diagrams in the cases shown in Fig.~\ref{fig:Figure3}(a) and Fig.~\ref{fig:Figure1app}(a), by employing the QEO and TEO methods, respectively. To carry out the numerics we use a Lenovo ThinkPad P15 workstation. We first consider the numerical evaluation for a grid mesh $10\times 10=100$ in the $(\beta_1/\omega,\beta)$ plane. For a discretization step $M=20$, we find that the required runtimes for the QEO and TEO methods are cor\-respon\-din\-gly $161$ and $1422$ seconds. For completeness, we also mention that the calculations of the abovementioned phase diagrams for the intermediate va\-lues of the step $M=\{5,10,15,20,25,30\}$ require the approximate runtimes of $\{337,684,
1075,1422,1775,2117\}$ seconds.

When instead considering a grid mesh $100\times 100=10000$ in the $(\beta_1/\omega,\beta)$ plane, which was actually employed to obtain the results shown in Fig.~\ref{fig:Figure3}(a) and Fig.~\ref{fig:Figure1app}(a) for a discretization step $M=20$, we find that the QEO and TEO runtimes become $15582$ and $129888$ seconds, respectively. By comparing the results for the two grids, we observe that the runtime increase by two orders of magnitude for each method closely follows the similar increase of the number of points of the grid mesh. Aside from the above conclusion, we are also in a position to assert that the computational runtime required for the TEO method is an order of magnitude larger than the one required when employing the QEO approach.

At this stage, we proceed by numerically inferring the minimum value of the discretization step $M$ that is required for the TEO and QEO methods to converge. Since, as discussed above, the runtime for a single phase diagram point progressively increases upon increasing the value for the discretization step $M$, in the remainder we simply restrict our investigation to a single point. In particular, we examine when the winding number value obtained for the point $(\beta_1/\omega,\beta)=(3,1)$ in Fig.~\ref{fig:Figure1app}(a) changes to the value obtained Fig.~\ref{fig:Figure3}(a). From our numerics, we find that TEO and QEO methods agree for quite a large value for the discretization step, i.e., above the minimum $M_{\rm min}=236$. Notably, the required runtime for this single point using the TEO method is about $\sim1420$ se\-conds. Hence, our above analysis further verifies the computational advantage of the QEO method for harmonically driven systems.}

\section{Stroboscopic effective Hamiltonian}\label{sec:BandInvariants}

We now complete the study of the TEO approach, by here considering the stroboscopic or effective Hamiltonian of the driven system. In this approach, one computes ${\cal U}(t,k)$ over a single period $T$ and then determines the effective Hamiltonian through $H_F(k)=(i/T)\ln\big[{\cal U}(T,k)\big]$. For systems with a DCS, as in the present case, $H_F(k)$ is block off-diagonal in the canonical basis. Hence, one can also define winding numbers using the effective Hamiltonian which are customarily termed band invariants $\nu_B$, since these neglect the micromotion of the system. As a result,  the band invariant depends on the chosen time frame~\cite{Asboth2014}. For instance, consider a new time frame in which the effective Hamiltonian preserves DCS. This new frame is obtained by shifting the time dependent Hamiltonian by $T/2$, and leads once again to the Hamiltonian of Eq.~\eqref{eq:HamPrime}, albeit now with $\theta_1(t,k)\mapsto-\theta_1(t,k)$. This yields a distinct band invariant $\bar{\nu}_B$. To clarify this, in Fig.~\ref{fig:Figure2app} we show our numerical results by obtaining the effective Hamiltonian for Eq.~\eqref{eq:HamPrime}, while using the parameter values of Fig.~\ref{fig:Figure3}(c)-(d). From Fig.~\ref{fig:Figure2app}  we find the relations $\nu_B=\nu_0+\nu_\pi$ and $\bar{\nu}_B=\nu_0-\nu_\pi$, which agree with the predictions of Ref.~\onlinecite{Asboth2014} for general chiral-symmetric driven systems.


\begin{thebibliography}{99}
\bibitem{Comment} {\color{black}We note that, strictly speaking, the SSH model belongs to class BDI, i.e., it exhibits a charge-conjugation symmetry. However, this symmetry can be broken, in contrast to the charge-conjugation symmetry encountered for a Kitaev chain. This difference arises because the SSH model was originally employed to describe polyacetylene and not a superconductor. From a topological classification point of view, one dimensional systems in either AIII or BDI classes are described using a $\mathbb{Z}$ topological inva\-riant. Therefore, there is no qualitative difference regar\-ding this aspect. However, here we choose to classify the SSH model by accounting for the fragility of its charge-conjugation symmetry in order to properly reflect that the topological excitations of the open SSH chain are electrons with a fractional charge and not neutral Majorana excitations. The latter strictly appear in D, DIII, and BDI classes which possess a robust charge-conjugation symmetry, that is built-in when adopting the Bogo\-liubov - de Gennes description. On the other hand, when it comes to the classification of the Kitaev chain model, we note that by restricting to a nearest-neighbor pairing term and by assuming an infinite uniform system, the superconducting phase factor becomes spatially independent and can be gauged away, thus, enlisting the system in the BDI class.}
\end{thebibliography}
\end{document}